\documentclass[iop]{emulateapj}

\usepackage{natbib}
\usepackage{amsmath, amssymb}
\usepackage{times}
\usepackage{apjfonts, graphicx, graphics}
\usepackage[breaklinks,colorlinks,urlcolor=blue,citecolor=blue]{hyperref}
\usepackage[all]{hypcap}
\usepackage{aas_macros}

\newcommand{\Mpc}{\rm\; Mpc}
\newcommand{\kpc}{\rm\; kpc}
\newcommand{\pc}{\rm\; pc}
\newcommand{\km}{\rm\; km}
\newcommand{\m}{\rm\; m}
\newcommand{\cm}{\rm\; cm}
\newcommand{\mum}{\hbox{$\rm\; \mu m\,$}}

\newcommand{\cmsq}{\hbox{$\cm^2\,$}}

\newcommand{\yr}{\rm\; yr}
\newcommand{\Gyr}{\rm\; Gyr}

\newcommand{\s}{\rm\; s}

\newcommand{\ks}{\rm\; ks}

\newcommand{\GHz}{\rm\; GHz}

\newcommand{\K}{\rm\; K}

\newcommand{\Msun}{\hbox{$\rm\thinspace M_{\odot}$}}

\newcommand{\Msunpyr}{\hbox{$\Msun\yr^{-1}\,$}}

\newcommand{\keV}{\rm\; keV}

\newcommand{\erg}{\rm\; erg}
\newcommand{\Jy}{\rm\; Jy}
\newcommand{\mJy}{\rm\; mJy}

\newcommand{\ergps}{\hbox{$\erg\s^{-1}\,$}}

\newcommand{\kmps}{\hbox{$\km\s^{-1}\,$}}

\newcommand{\kmpspMpc}{\hbox{$\kmps\Mpc^{-1}\,$}}

\newcommand{\Lsun}{\hbox{$\rm\thinspace L_{\odot}$}}

\newcommand{\pcmsq}{\hbox{$\cm^{-2}\,$}}

\providecommand{\e}[1]{\ensuremath{\times 10^{#1}}}

\newcommand{\Jykmps}{\hbox{$\Jy\km\s^{-1}\,$}}

\newcommand{\Kkmpspcsq}{\hbox{$\K\km\s^{-1}\pc^{2}\,$}}
\newcommand{\acounits}{\hbox{$\Msun(\Kkmpspcsq)^{-1}$}}
\newcommand{\uJy}{\rm\; \mu Jy}

\shorttitle{Molecular Gas in RXJ0821}
\shortauthors{Vantyghem et al.}

\begin{document}

\title{An enormous molecular gas flow in the RXJ0821+0752 galaxy cluster}
\author{A.~N. Vantyghem$^{1, 2}$}
\author{B.~R. McNamara$^{1, 3}$}
\author{H.~R. Russell$^{4}$}
\author{A.~C. Edge$^5$}
\author{P.~E.~J. Nulsen$^{6, 7}$}
\author{F. Combes$^{8, 9}$}
\author{A.~C. Fabian$^{4}$}
\author{M. McDonald$^{10}$}
\author{P. Salom{\'e}$^{8}$}

\affil{
    $^1$ Department of Physics and Astronomy, University of Waterloo, Waterloo, ON N2L 3G1, Canada; \href{mailto:a2vantyg@uwaterloo.ca}{a2vantyg@uwaterloo.ca} \\ 
	$^2$ University of Manitoba, Department of Physics and Astronomy, Winnipeg, MB R3T 2N2, Canada \\
    $^3$ Perimeter Institute for Theoretical Physics, Waterloo, Canada \\
    $^4$ Institute of Astronomy, Madingley Road, Cambridge CB3 0HA \\
    $^5$ Department of Physics, Durham University, Durham DH1 3LE \\
    $^6$ Harvard-Smithsonian Center for Astrophysics, 60 Garden Street, Cambridge, MA 02138, USA \\
    $^7$ ICRAR, University of Western Australia, 35 Stirling Hwy, Crawley, WA 6009, Australia \\
    $^8$ LERMA, Observatoire de Paris, PSL Research Univ., Coll{\`e}ge de France, CNRS, Sorbonne Univ., UPMC, Paris, France \\
    $^9$ Coll{\`e}ge de France, 11 place Marcelin Berthelot, 75005 Paris \\
    $^{10}$ Kavli Institute for Astrophysics and Space Research, Massachusetts Institute of Technology, 77 Massachusetts Avenue, Cambridge, MA 02139, USA \\
}

\begin{abstract}

We present recent {\it Chandra} X-ray observations of the RXJ0821.0+0752 galaxy cluster in addition to ALMA observations of the CO(1-0) and CO(3-2) line emission tracing the molecular gas in its central galaxy. All of the CO line emission, originating from a $10^{10}\,M_{\odot}$ molecular gas reservoir, is located several kpc away from the nucleus of the central galaxy. The cold gas is concentrated into two main clumps surrounded by a diffuse envelope. They form a wide filament coincident with a plume of bright X-ray emission emanating from the cluster core. This plume encompasses a putative X-ray cavity that is only large enough to have uplifted a few percent of the molecular gas. Unlike other brightest cluster galaxies, stimulated cooling, where X-ray cavities lift low entropy cluster gas until it becomes thermally unstable, cannot have produced the observed gas reservoir. Instead, the molecular gas has likely formed as a result of sloshing motions in the intracluster medium induced by a nearby galaxy. Sloshing can emulate uplift by dislodging gas from the galactic center. This gas has the shortest cooling time, so will condense if disrupted for long enough.

\end{abstract}



\section{Introduction}

Located at the centers of galaxy clusters, brightest cluster galaxies (BCGs) are the most massive and luminous galaxies known. They are giant ellipticals with extended, diffuse stellar envelopes and stellar populations that are primarily old and dormant. However, those located in cool core clusters, where the central atmospheric cooling time falls below $\sim1\Gyr$, are replete with cold gas and star formation. Their molecular gas masses, which can exceed $10^{10}\Msun$, surpass those of gas-rich spirals \citep{Edge01, Edge02, Edge03, Salome03}. Star formation proceeding at rates of several to several hundred solar masses per year \citep[e.g.][]{mcn04, ODea08, McDonald11, Donahue15, Tremblay15, McDonald18} place BCGs among starbursts on the Kennicutt-Schmidt relation \citep{Kennicutt98, Kennicutt12}. 

An abundance of observational evidence indicates that this cold gas and star formation originates from the condensation of the hot intracluster medium (ICM). Molecular gas is regularly associated with filamentary emission observed in H$\alpha$ \citep[e.g.][]{Lynds70, Heckman81, Cowie83, Hu85, Crawford99} and soft X-rays \citep[e.g.][]{Fabian01, Fabian03, Werner13, Walker15}, implying that BCGs contain multiphase gas spanning five decades in temperature. Moreover, molecular gas, nebular, emission, and star formation are observed preferentially when the central atmospheric cooling time falls below $\sim5\e{8}\yr$, or equivalently when the entropy falls below $30\keV\cmsq$ \citep{Cavagnolo08, Rafferty08, Hogan17b, Pulido18}. Correlations between the rates of star formation and mass deposition from the ICM further support this picture \citep{Egami06, ODea08}.

Uninhibited cooling would result in hundreds to thousands of solar masses per year of gas condensing out of the ICM. Despite the wealth of evidence that the ICM is condensing, cooling ensues at only a few percent of the expected rate \citep{Peterson06}. Instead, active galactic nucleus (AGN) feedback injects heat into the surrounding atmosphere, regulating the rate of cooling \citep[for reviews, see][]{McNamara07, McNamara12, Fabian12}. In the ``radio-mode'' mechanical feedback that operates in giant ellipticals and galaxy clusters, radio jets launched by the central AGN inflate bubbles (X-ray cavities), drive shock fronts, and generate sound waves in the hot atmosphere \citep[e.g.][]{mcn00, Blanton01, Fabian06}. The power output by the AGN is closely coupled to the cooling rate \citep{Birzan04, Dunn06, Rafferty06}, implying that AGN can regulate the growth of their host galaxy over long timescales. AGN feedback is fueled through black hole accretion, likely of the molecular gas that has condensed from the hot atmosphere \citep{Pizzolato05, Gaspari13, Li14a}. This establishes a feedback loop, wherein the ICM cools and condenses into the cold gas that accretes onto the nuclear black hole and fuels the energetic outbursts that reheat the surrounding hot phase.

While AGN feedback primarily affects the volume-filling hot atmosphere, recent observations indicate that it couples to the dense molecular phase as well. Fast, jet-driven outflows of ionized and molecular gas have been detected in radio galaxies \citep{Morganti05, Nesvadba06, Alatalo11, Dasyra11, Tadhunter14, Morganti15}. Direct interactions between radio jets and molecular gas have also been observed in BCGs. The molecular gas in M87 is located at the truncation of the radio lobe, appearing to be either excited or destroyed by AGN activity \citep{Simionescu18}. Jet-induced star formation is also observed in Centaurus A \citep{Salome17}. In A1795 the radio jet is projected along the inner edge of a curved molecular filament, suggesting that the jet has either deflected off of the molecular gas or the gas is entrained in the expanding radio bubble \citep{Russell17b}.

A broader consensus is emerging that the formation of cold gas in BCGs is also stimulated by AGN activity. The molecular filaments identified with either ALMA observations or through their nebular emission extend radially away from the galactic center and frequently trail X-ray cavities \citep[e.g.][]{Conselice01, Hatch06, Salome06, Salome11, Lim12, McDonald12, mcn14, Russell14, Russell16, Russell17, Vantyghem16, Vantyghem18}. Either molecular clouds are lifted directly by X-ray cavities or they have condensed from thermally unstable, low entropy gas originally in the cluster core that has been lifted by the cavity \citep{Revaz08, mcn16}. The shallow velocity gradients along the filaments suggest that the molecular clouds are supported against freefall, and are potentially pinned to the hot atmosphere via magnetic fields \citep{Fabian08, Russell16, Russell17, Vantyghem16, Vantyghem18}. Moreover, the velocities are well below the stellar velocity dispersion, so are too slow to escape the central galaxy in an outflow. Indeed, redshifted absorption lines imply that clouds are returning in a circulating flow and accreting onto the central supermassive black hole \citep{David14, Tremblay16}.


Here we present a multi-wavelength analysis of the RXJ0821+0752 (hereafter RXJ0821) galaxy cluster. We present ALMA observations of the CO(1-0) and CO(3-2) rotational emission lines tracing the molecular gas in the central galaxy alongside a new $63.5\ks$ {\it Chandra} X-ray observation. These ALMA observations were first presented in \citet[][hereafter V17]{Vantyghem17}, which used the intensities of the emission lines, including the $^{13}$CO(3-2) line, to estimate the CO-to-H$_2$ conversion factor for the first time in a BCG. In this work we focus on the morphology and kinematics of the molecular gas, relating it to features in the X-ray image.

RXJ0821 contains one of the most gas-rich BCGs known. A strong CO detection from the IRAM 30-m telescope implied a molecular gas mass of $2\e{10}\Msun$ \citep[][corrected for cosmology]{Edge01}. Follow-up observations with the OVRO interferometer marginally resolved the cold gas, showing an extension west of the BCG \citep{Edge03}. Emission from the $1-0$ S series lines of H$_2$ was detected in the BCG but not in the western extension \citep{Edge02}. 
The central galaxy also hosts a luminous emission-line nebula \citep[$L_{{\rm H}\alpha} = 2.55\e{42}\ergps$ --][]{BayerKim02, Hatch07} and an infrared luminosity, $L_{\rm IR}=8.47\e{44}\ergps$, that implies a star formation rate of $37\Msunpyr$ \citep{Quillen08, ODea08}. Unlike many other cool core clusters, the radio source in RXJ0821 is exceptionally weak. At $5\GHz$ the flux density is $0.85\pm0.07\mJy$, making it the third weakest radio source in the Brightest Cluster Survey \citep{BayerKim02, Hogan15a}. The radio source is also offset from the BCG nucleus by $2.7\kpc$. RXJ0821 may be undergoing an evolutionary phase dominated by the cooling flow. 

Throughout this work we assume a standard $\Lambda$-CDM cosmology with $H_0=70\kmpspMpc$, $\Omega_{{\rm m}, 0}=0.3$, and $\Omega_{\Lambda, 0}=0.7$. At the redshift of RXJ0821 ($z=0.11087$; see Section \ref{sec:systemic}), the angular scale is $1''=2.0\kpc$ and the luminosity distance is $510\Mpc$.

\section{Observations and Data Reduction}

\subsection{Chandra}

RXJ0821.0+0752 was observed for 29~ks on 2014-12-15 (ObsID 17194) and 37~ks on 2014-12-28 (ObsID 17563) using the ACIS-S3 detector on the {\it Chandra X-ray Observatory}. The observations were reprocessed using {\sc ciao} version 4.5 and {\sc caldb} version 4.6.7. We applied charge transfer inefficiency and time-dependent gain corrections to the level 1 event files, which were then filtered to remove photons with bad grades. Periods affected by flares were identified and filtered using the {\sc lc\_clean} script. The final exposure time of the cleaned data was 63.5~ks.

\begin{figure*}
	\centering
	\includegraphics[width=0.32\textwidth]{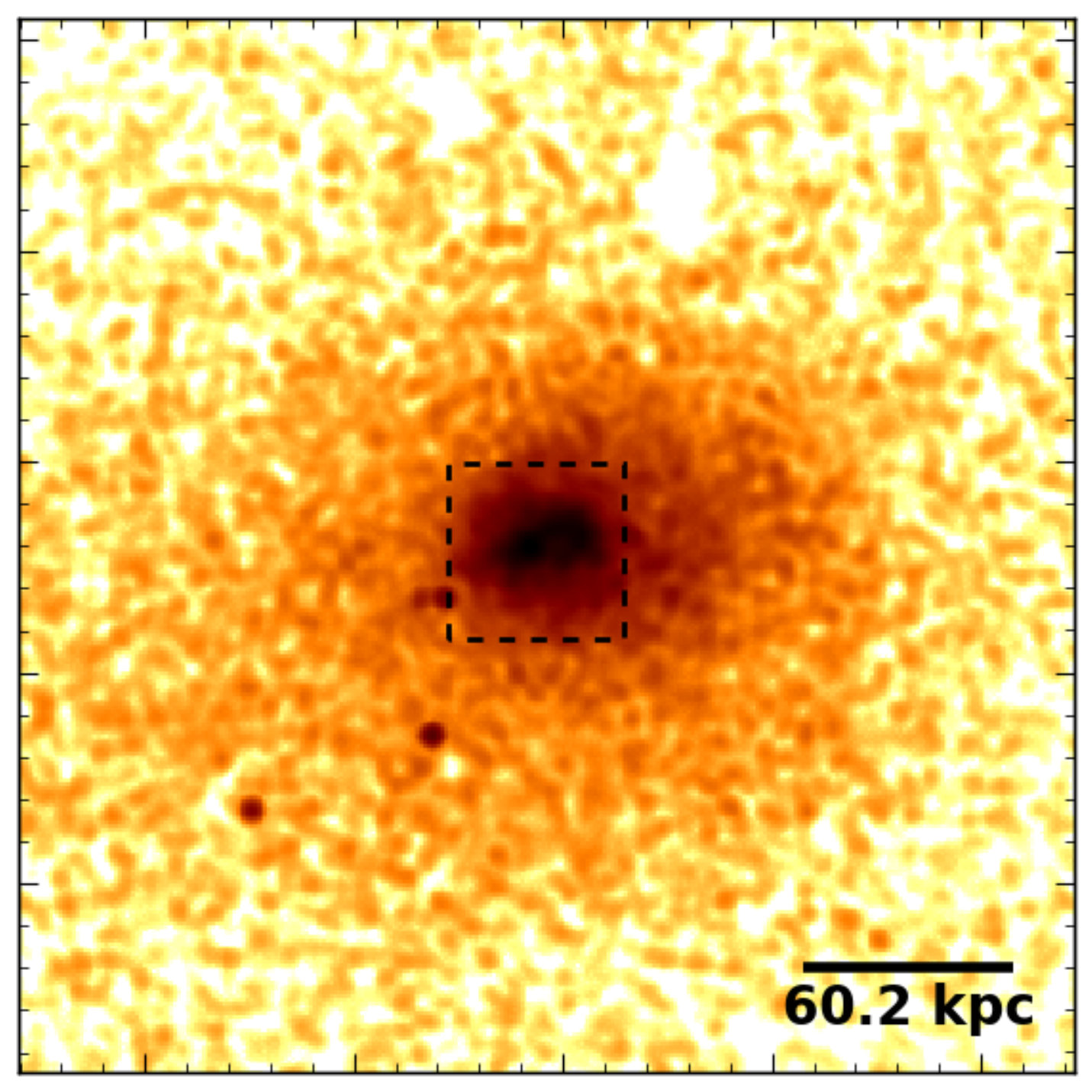}
	\includegraphics[width=0.32\textwidth]{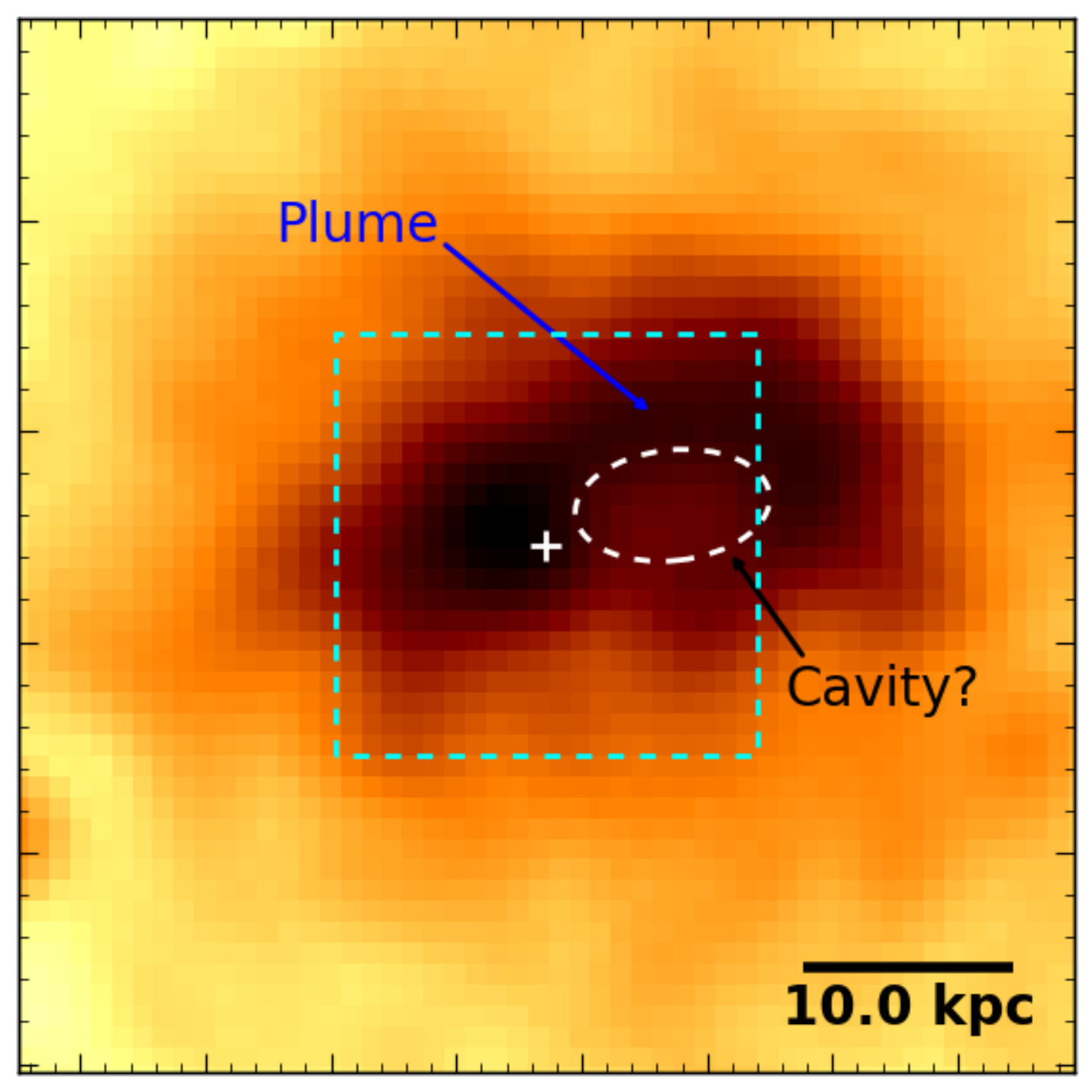}
	\includegraphics[width=0.32\textwidth]{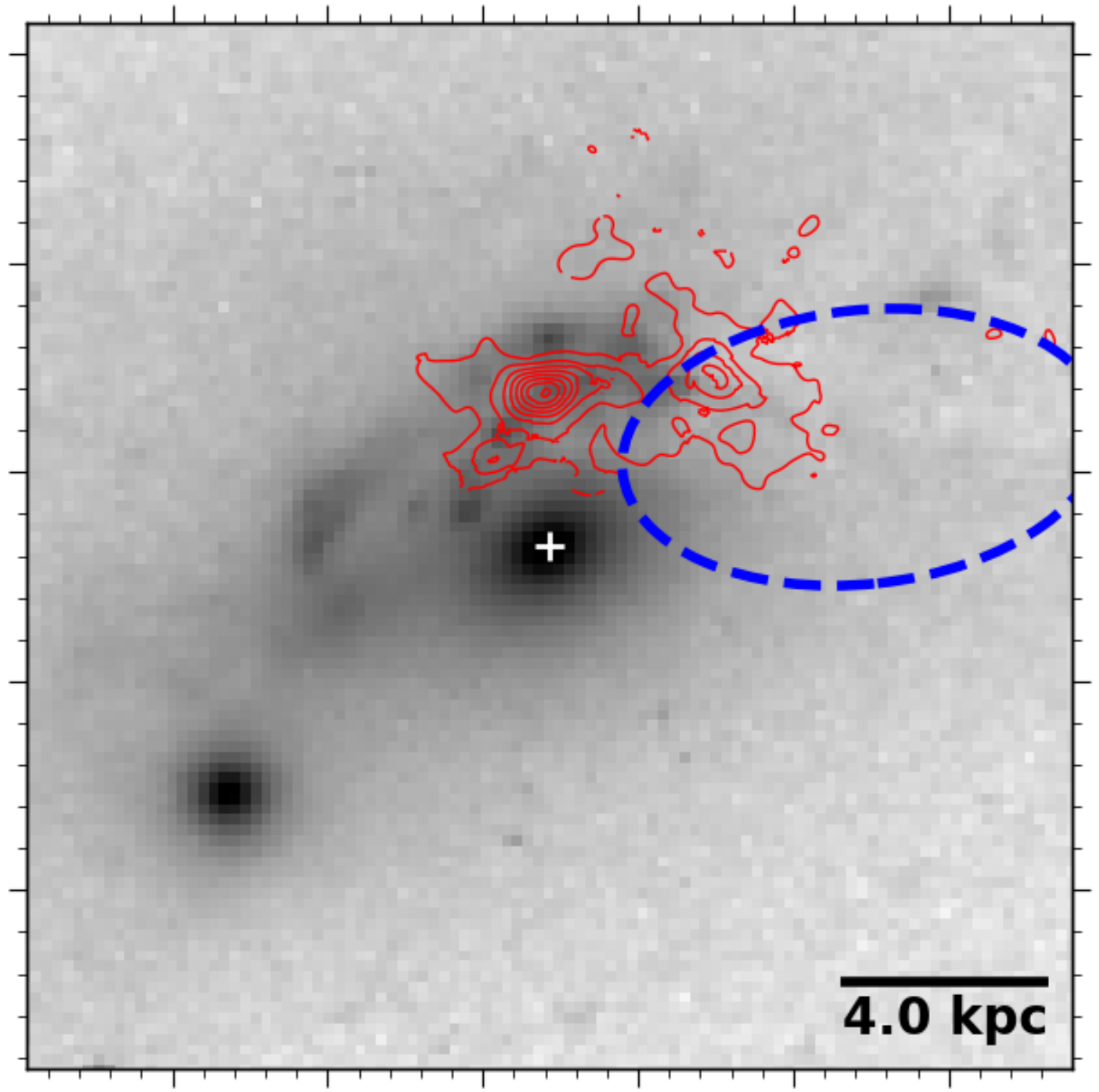}
	\caption[X-ray and optical images of RXJ0821.0+0752]{
      Background-subtracted {\it Chandra} X-ray surface brightness images (left and center) and an HST F606W optical image (right) with contours of the CO(3-2) emission overlaid. The dashed ellipse indicates the location of the putative X-ray cavity (see Section \ref{sec:cavity}). The cross indicates the location of the BCG nucleus.
    }
	\label{fig:xray}
\end{figure*}

The final background-subtracted image, shown in Fig. \ref{fig:xray} (left), was created by reprojecting the observations to match the position of the longest exposure (ObsID 17563) and summing all events in the $0.5-7\keV$ energy range. Point sources were identified and removed using {\sc wavdetect} and confirmed via visual inspection. 
Blank-sky backgrounds were extracted for each observation, processed the same way as the events files, and reprojected to the corresponding position on the sky. The blank-sky backgrounds were normalized to match the observed count rate in the $9.5-12\keV$ energy range.

\subsection{ALMA}

The ALMA observations of RXJ0821+0752 are described in detail in V17. Briefly, the observations targeted the CO(1-0) and CO(3-2) lines, which, located at redshifted frequencies of $103.848\GHz$ and $311.528\GHz$, fell in Bands 3 and 7, respectively. The observations (Cycle 4, ID 2016.1.01269.S, PI McNamara) were conducted on 2016 Oct 30 (Band 3; 86.7 minutes on source) and Nov 4 and 2016 Oct 1 (Band 7; 22.7 minutes on source). An additional baseband in the Band 7 observation also covered the $^{13}$CO(3-2) line at $297.827\GHz$. The remaining three Band 3 basebands and two Band 7 basebands were used to measure the sub-mm continuum emission. Both observations employed 40 antennas, with baselines ranging from $18-1124\m$ for Band 3 and $15-3247\m$ for Band 7.

The observations were calibrated in {\sc casa} version 4.7.0 \citep{casa} using the pipeline reduction scripts. Continuum-subtracted data cubes were created using {\sc uvcontsub} and {\sc clean}. Images of the line emission were reconstructed using Briggs weighting with a robust parameter of 2. An additional $uv$ tapering was used to smooth the CO(3-2) image on scales below 0.1 arcsec. The final CO(1-0) and CO(3-2) data cubes had synthesized beams of $0.61''\times 0.59''$ (P.A. $-70.4^{\circ}$) and $0.21''\times 0.165''$ (P.A. $37.2^{\circ}$), respectively. The CO(1-0) and CO(3-2) images were binned to $3$ and $5\kmps$ velocity channels, respectively. The RMS noise in the line-free channels were $0.5$ and $1.1\mJy~{\rm beam}^{-1}$, respectively.

\subsubsection{Systemic Velocity}
\label{sec:systemic}

Throughout this work we adopt a systemic redshift of $z=0.11087\pm0.00004$, measured using the spatially integrated ALMA CO spectra. This is consistent with longslit \citep[$z=0.110\pm0.001$;][]{Crawford95} and {\it VIMOS} integral field unit \citep[$z=0.11088$;][]{Hamer16} observations of the H$\alpha + [{\rm N}\,\textsc{ii}]$ complex. As each of these measurements probes either the molecular or ionized gas within the BCG, they do not necessarily reflect the systemic stellar velocity. \citet{Crawford95} also detected stellar absorption lines, reporting the same redshift as for the emission lines. However, the high uncertainties ($300\kmps$) prohibit us from using this value for the systemic velocity. An accurate measurement of the relative velocity between the gas and stars is crucial in understanding gas flows. Since an accurate stellar velocity is absent, we measure our velocities in the rest frame of the gas and interpret any gas flows cautiously.

\section{Cluster X-ray Properties}

In Fig. 1 we present the $0.5-7\keV$ band {\it Chandra} X-ray image. On large scales the X-ray surface brightness is relatively smooth. No surface brightness edges indicative of shocks or cold fronts are evident in the image, although they may be present but not detected. 
Within the central $20\kpc$ the surface brightness distribution is more complex. An arc extends NW of the cluster center, curving southward after $\sim12\kpc$. The elliptical surface brightness depression encompassed by this arc may correspond to an X-ray cavity. This is discussed further in Section \ref{sec:cavity}.

\subsection{Radial Profiles of Gas Properties}
\label{sec:profiles}

Spectra were extracted from eight annular regions extending out to $600\kpc$. Each region had a minimum of 2500 net counts. The annuli were centered on the BCG nucleus as determined from the HST F606W image. The optical centroid is located $2\kpc$ SW of the peak X-ray flux. At radii of $\sim20\kpc$ the X-ray emission is better centered $3.3\kpc$ NW of the optical centroid. The radial profiles are only weakly affected by which of these three centroids is used. Since the cavity (see Section \ref{sec:cavity}) is located close to the center, the cavity age depends strongly on the adopted centroid. The optical centroid was chosen because it is the median of the centroids, best reflects the gravitating mass near the cluster center, and was used to measure the mass profile \citep{Hogan17b}.

The spectroscopic analysis of the X-ray data was performed using {\sc xspec} v12.7.1 \citep{Arnaud96}. Spectral deprojection was performed using the geometric method {\sc dsdeproj} \citep{Sanders07, Russell08}. This removes the spectral contribution of gas along the line-of-sight projected into an annulus. The projected and deprojected spectra were both fit by a single temperature thermal model with photoelectric absorption, {\sc phabs}$\times${\sc apec}. The foreground hydrogen column density was fixed to the Galactic value of $N_H = 2.01\e{20}\pcmsq$ \citep{Kalberla05}. Temperature, normalization, and metallicity were all allowed to vary, with the metal abundance ratios taken from \citet{Anders89}. The projected spectra were left unbinned and fit with C-statistics, while the deprojected spectra were grouped to a minimum of 25 counts per energy bin and fit with the $\chi^2$ statistic.

\begin{figure*}
  \begin{minipage}{\textwidth}
    \centering
    \includegraphics[width=0.32\textwidth]{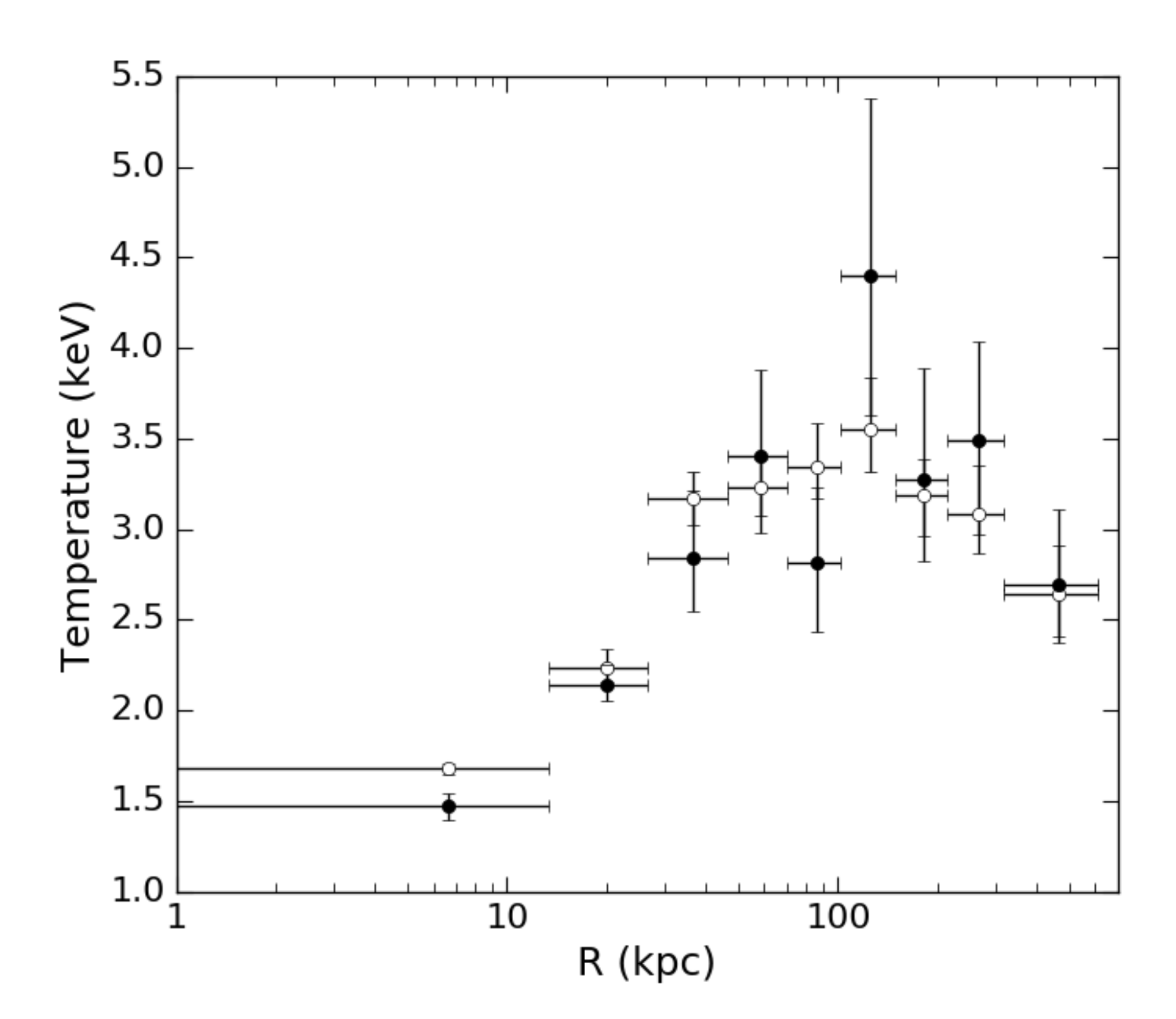}
    \includegraphics[width=0.32\textwidth]{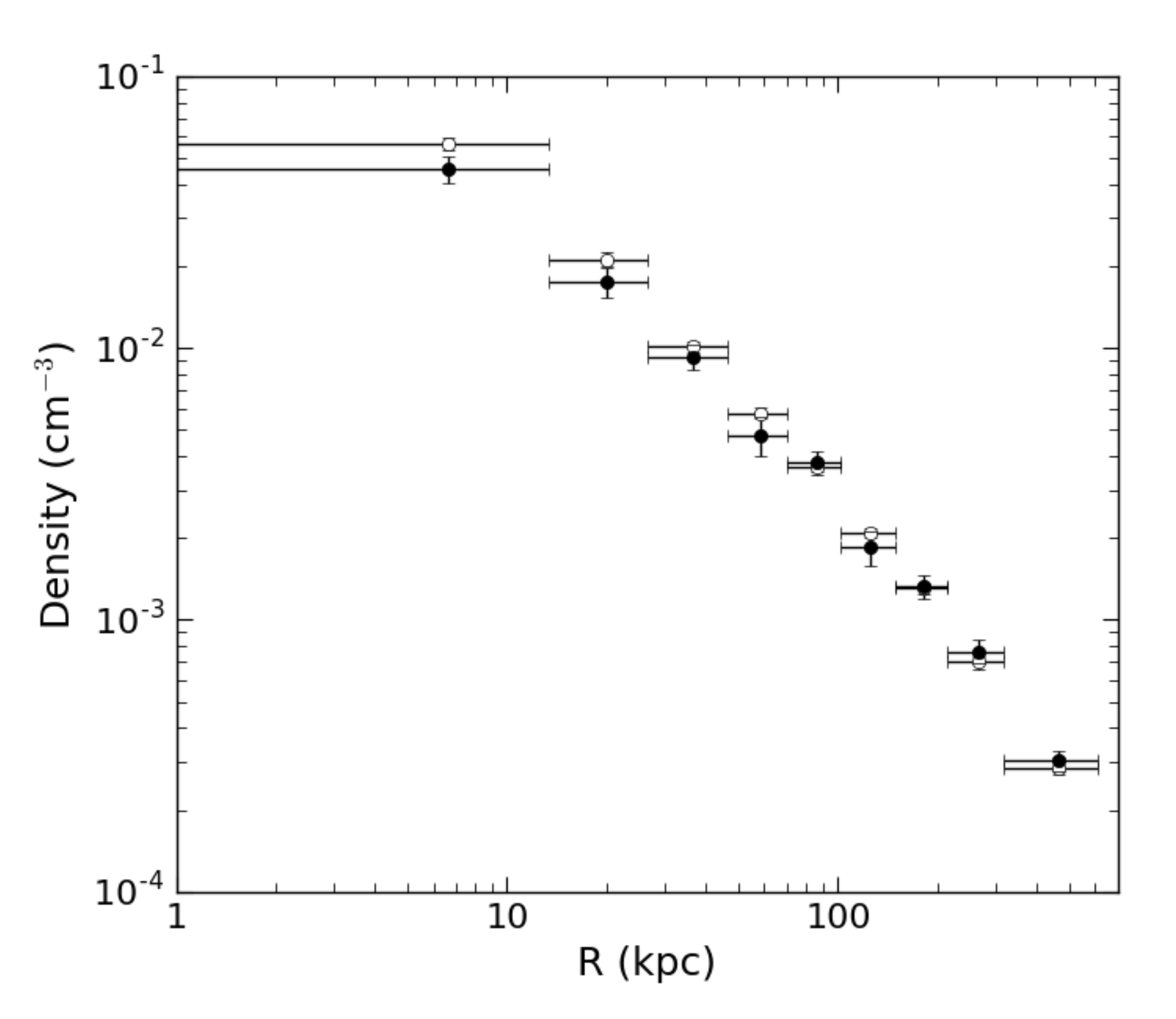}
    \includegraphics[width=0.32\textwidth]{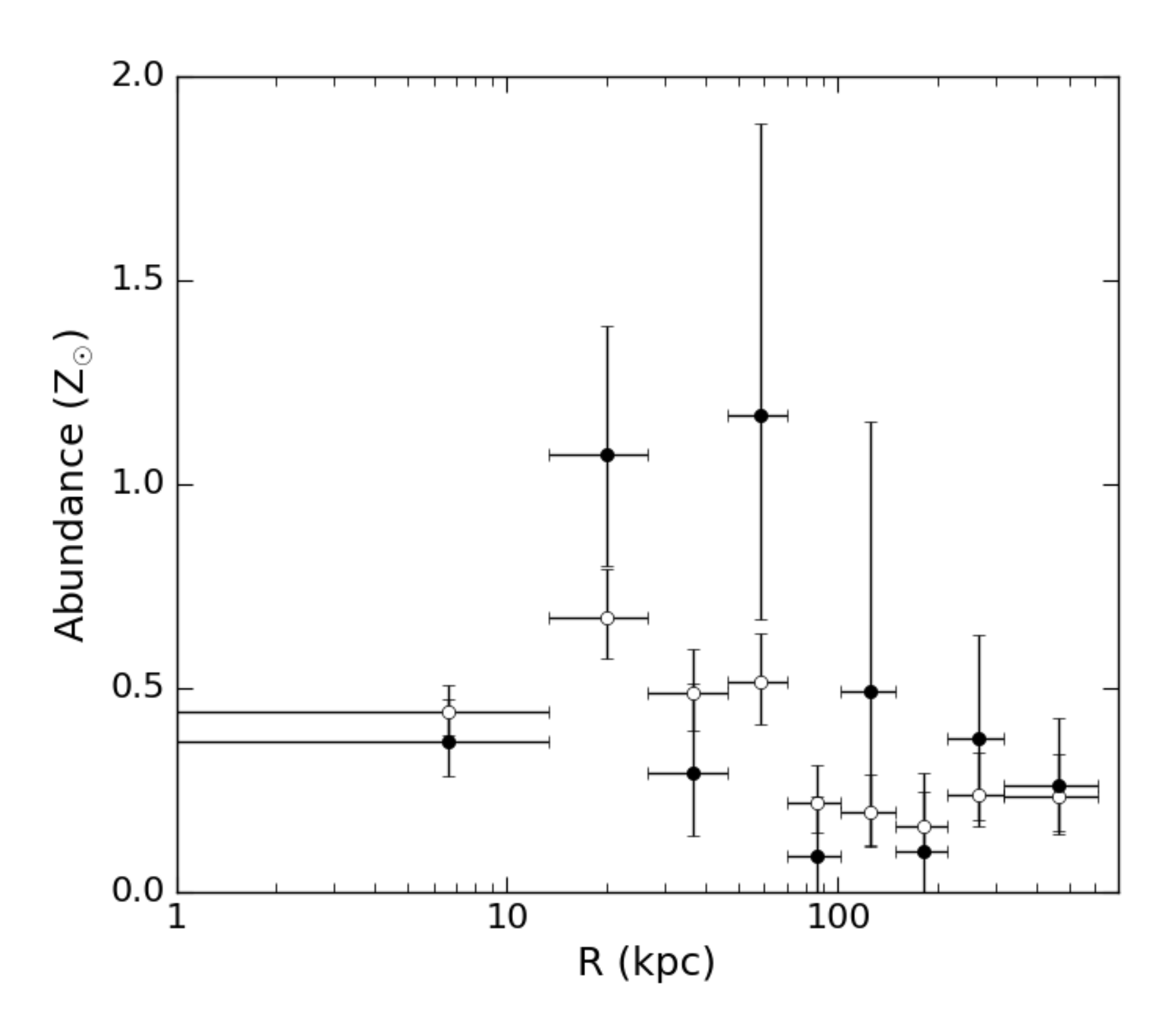}
  \end{minipage}
  \begin{minipage}{\textwidth}
    \centering
    \includegraphics[width=0.32\textwidth]{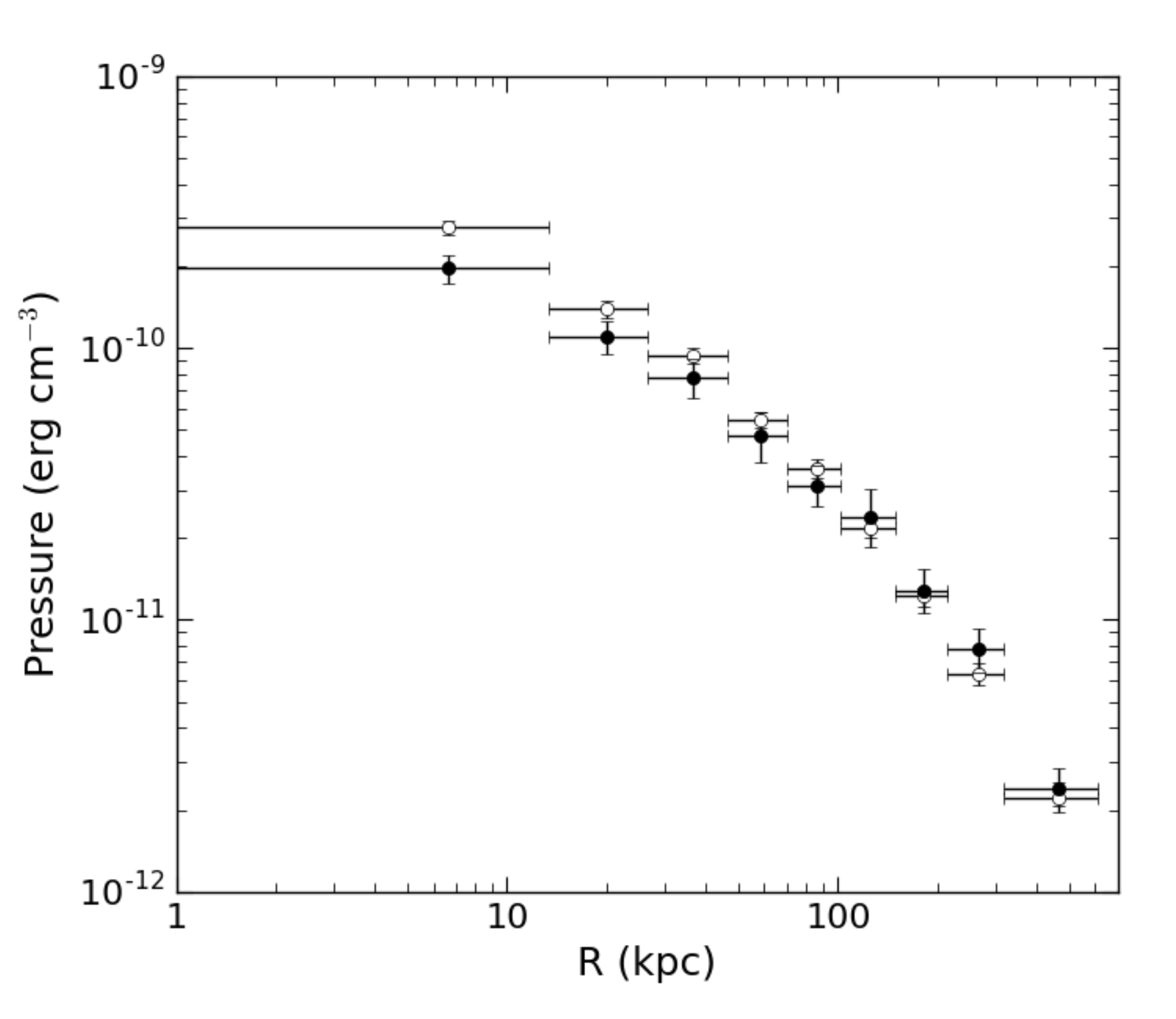}
    \includegraphics[width=0.32\textwidth]{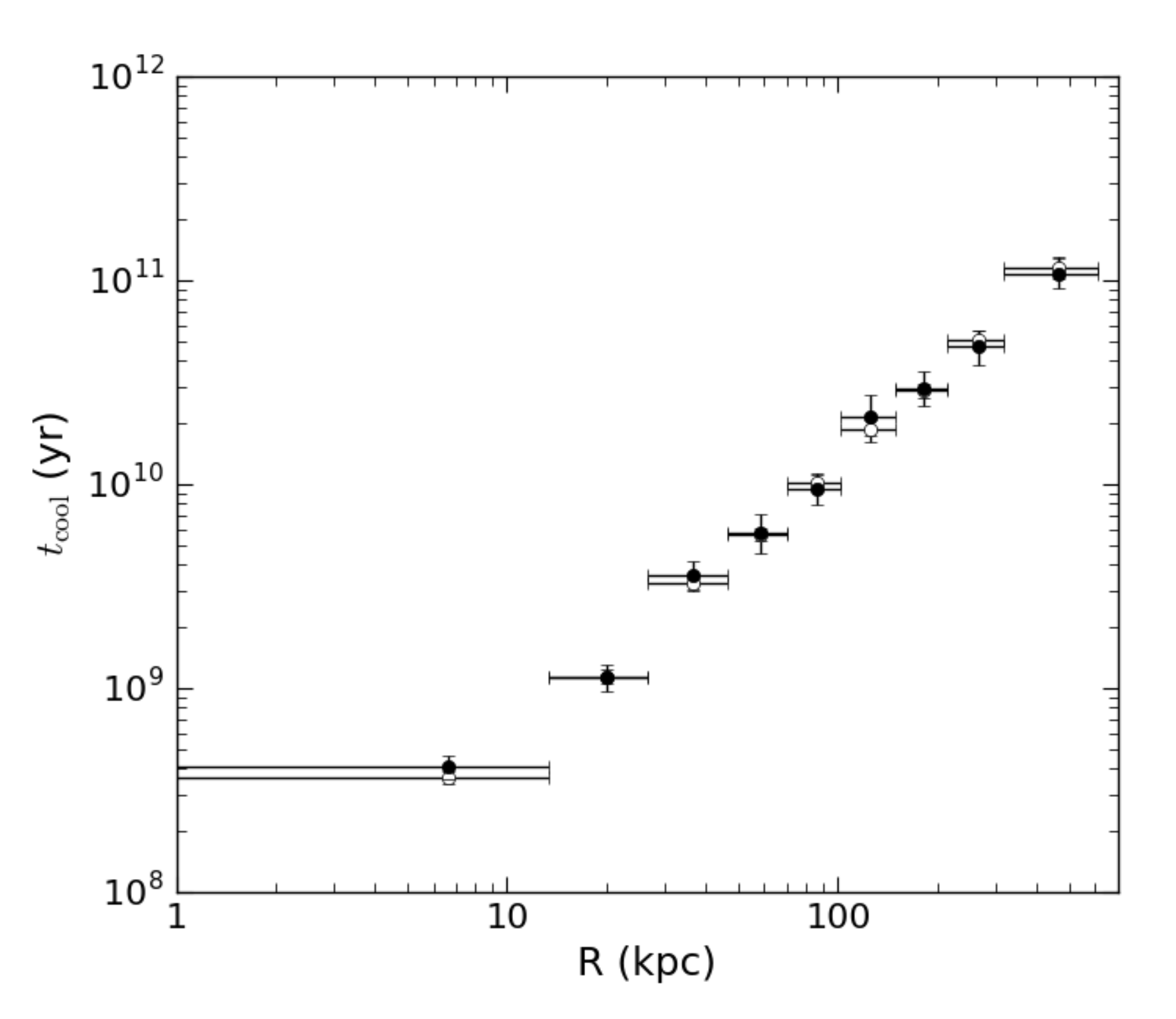}
    \includegraphics[width=0.32\textwidth]{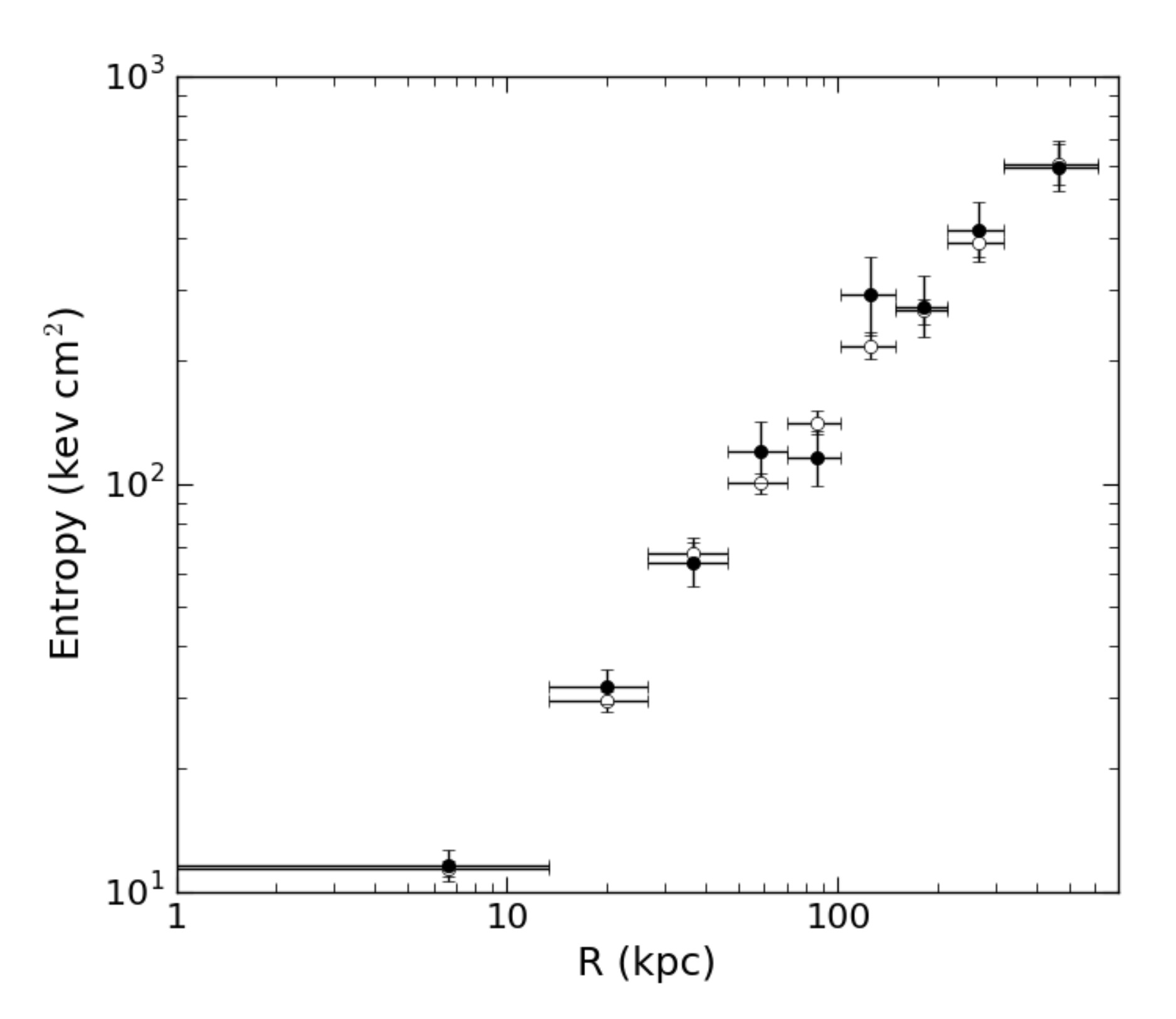}
  \end{minipage}
  \caption[X-ray derived profiles of RXJ0821]{
    Projected (open circles) and deprojected (filled circles) profiles of X-ray derived quantities.
  }
  \label{fig:profiles}
\end{figure*}

The normalization of the {\sc apec} model is related to the gas density via
\begin{equation}
  {\rm norm} = \frac{10^{-14}}{4\pi [D_A~(1+z)]^2} \int n_e n_H {\rm d}V,
\end{equation}
where $D_A$ is the angular diameter distance, V is the volume of the annulus, and $n_H=n_e/1.2$ is assumed to be constant within each annulus. The gas pressure is determined from density and temperature using the ideal gas law, $p=(n_e+n_H)~kT=1.8n_e kT$. The entropy index of the gas is defined as $K=kT~n_e^{-2/3}$. The cooling time, which is the time it would take for the gas to radiate away all of its thermal energy, is given by
\begin{equation}
  t_{\rm cool} = \frac{3}{2} \frac{p}{n_e n_H \Lambda(T, Z)}.
\end{equation}
The cooling function, $\Lambda(T, Z)$, was determined from the bolometric X-ray luminosity, $L_x=\int n_e n_H \Lambda(T, Z) {\rm d}V$, which was obtained by integrating the unabsorbed thermal model between $0.1$ and $100\keV$. 
The projected and deprojected profiles for each of these quantities is shown in Fig. \ref{fig:profiles}.

\subsection{The X-ray ``Cavity''}
\label{sec:cavity}

The X-ray image shown in Fig. \ref{fig:xray} (center) shows a surface brightness depression located $6\kpc$ NW of the BCG nucleus. The north side of the surface brightness depression is encompassed by an X-ray bright filament. Without an image of the radio source we cannot conclusively determine if this feature is an X-ray cavity. The existing radio data shows no extended structure on $2-5$~arcsec scales at $1.4$ or $5\GHz$, although the radio emission is offset along the direction of the X-ray filament \citep{BayerKim02}. Our ALMA imaging shows that the continuum emission at $98.8$ and $304.6\GHz$ is similarly offset from the BCG nucleus and is coincident with the molecular gas. The lack of radio emission from the galactic center indicates that the AGN is currently quiescent, so the surface brightness depression may simply be a lack of bright emission when contrasted with the X-ray filament. 
Nevertheless, we measure the size of the surface brightness depression and compute its energetics should it be a real cavity. Throughout this work we refer to this feature as simply the X-ray cavity, though it should be taken with the caveat stated here.

The projected size of the X-ray cavity was determined by qualitatively fitting an ellipse to the surface brightness depression. The identified region is shown in Fig. \ref{fig:xray}. The enthalpy required to inflate the cavity is given by $E_{\rm cav}=4pV$, where the prefactor is suitable for a relativistic gas filling the cavity volume. The pressure within the cavity, $p$, was determined by assuming that the cavity is in pressure balance with its surroundings, taking the deprojected pressure at a projected radius equal to the distance to the cavity center. The cavity volume was computed by assuming that its size along the line-of-sight is given by the geometric mean of the semi-major and semi-minor axes, $r=\sqrt{ab}$, so that $V=\frac{4}{3}\pi(ab)^{3/2}$.

We estimate the age of the cavity using both the buoyant rise time and sound crossing time \citep{Birzan04}. The sound crossing time is simply a function of ICM temperature, and is given by
\begin{equation}
t_{c_s} = R/c_s = R\sqrt{\mu m_H / \gamma kT},
\end{equation} 
where $\gamma=5/3$ for an ideal gas. The buoyant rise time is the time taken for the cavity to rise to its current projected distance at its terminal velocity,
\begin{equation}
t_{\rm buoy} = R/v_t \sim R\sqrt{SC/2gV}.
\end{equation}
Here $C=0.75$ is the drag coefficient \citep{Churazov01} and $S$ is the bubble cross-section, which is assumed to be its projected area. The acceleration under gravity is determined using the cluster mass profiles from \citet{Hogan17b}. We find that the two timescales are comparable, and adopt the buoyancy time when computing the mean cavity power, $P_{\rm cav} = E_{\rm cav}/t_{\rm buoy}$.

The size and energetics of the cavity and corresponding AGN outburst are summarized in Table \ref{tab:cavity}. The power output by the AGN, $1.3\e{43}\ergps$, is low compared to other cool core clusters. It is ten times less powerful than the outburst in Perseus, which itself is only a moderately powerful system \citep{Rafferty06}. Fueling the outburst through accretion requires an accreted mass of $M_{\rm acc} = E_{\rm cav}/\epsilon c^2 = 2.4\e{4}\Msun$, assuming an efficiency of $\epsilon=0.1$. Although no nuclear molecular gas has been detected, the accreted mass can easily be supplied by even a small fraction of the total molecular gas supply, which totals $10^{10}\Msun$.

\begin{table}
\caption{Cavity Measurements}
\begin{center}
\begin{tabular}{l l}
\hline
\hline
$a$             & $4.7\kpc$ \\
$b$             & $2.6\kpc$ \\
$R$             & $6.4\kpc$ \\
$4pV$           & $4.2\e{57}\erg$ \\
$t_{\rm sc}$    & $10.1\e{6}\yr$ \\
$t_{\rm buoy}$  & $10.5\e{6}\yr$ \\
$P_{\rm cav}$   & $1.3\e{43}\ergps$ \\
$M_{\rm acc}$   & $2.4\e{4}\Msun$ \\
$M_{\rm displaced}$ & $2.3\e{8}\Msun$ \\
\hline
\hline
\end{tabular}
\end{center}
\label{tab:cavity}
\end{table}

\subsection{Cooling of the Hot Atmosphere}
\label{sec:cooling}

To determine the amount of gas that may be cooling out of the hot atmosphere, we added an {\sc mkcflow} component to the previous thermal model. This is a classical cooling flow model for gas at a constant pressure cooling through a specified temperature range. The maximum temperature was taken to be the mean temperature within that annulus, and the minimum temperature limit was set at $0.1\keV$. The mass deposition rate yielded by this model is therefore an upper limit on the amount of gas that cools below $0.1\keV$ and condenses out of the ICM. Abundances in the cooling flow component were tied to the thermal model.

Following \citet{McDonald10}, we define the cooling radius, $r_{\rm cool}$, as the radius where the cooling time falls below $5\Gyr$. From the profiles in Section \ref{sec:profiles} we obtain $r_{\rm cool}=50\kpc$. A spectrum was extracted from this region and deprojected using the spectra from a series of overlying annuli using the same method as Section \ref{sec:profiles}. The best-fitting mass deposition rate is $\dot{M}_{\rm cool} = 34\pm10\Msunpyr$, and the luminosity of this cooling gas is $3.8\e{42}\ergps$. The total X-ray luminosity within this region is $L_X=2.87\pm0.05\e{43}\ergps$, so heating must offset $>85\%$ of the radiative losses within the central $50\kpc$. This spectroscopic mass deposition rate is consistent with the measured star formation rate.

\section{Molecular Gas Properties}

Maps of integrated flux, velocity, and FWHM of the CO(1-0) and CO(3-2) lines were created by fitting the spectra extracted from individual pixels, averaged over a box the size of the synthesized beam, of the respective ALMA images. Up to two Gaussian components, representing multiple coincident velocity structures, were used to model each pixel's spectrum. The significance of each velocity component was tested using a Monte Carlo analysis with at least 2500 iterations. A detection required $3\sigma$ significance. The presence of one component was required before attempting to fit a second. Instrumental broadening has been incorporated into the model.

\begin{figure}
  \centering
  \includegraphics[width=\columnwidth]{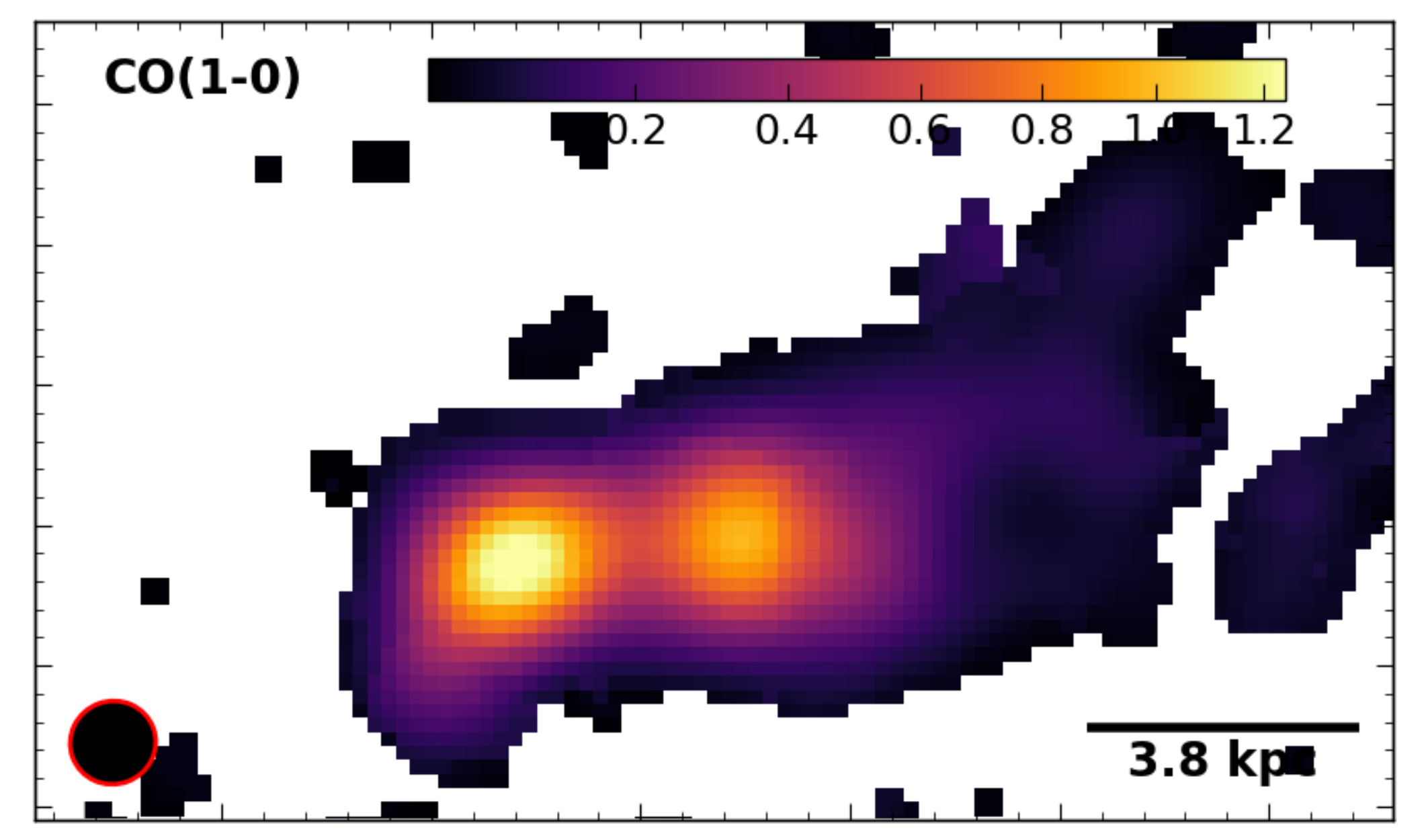}
  \includegraphics[width=\columnwidth]{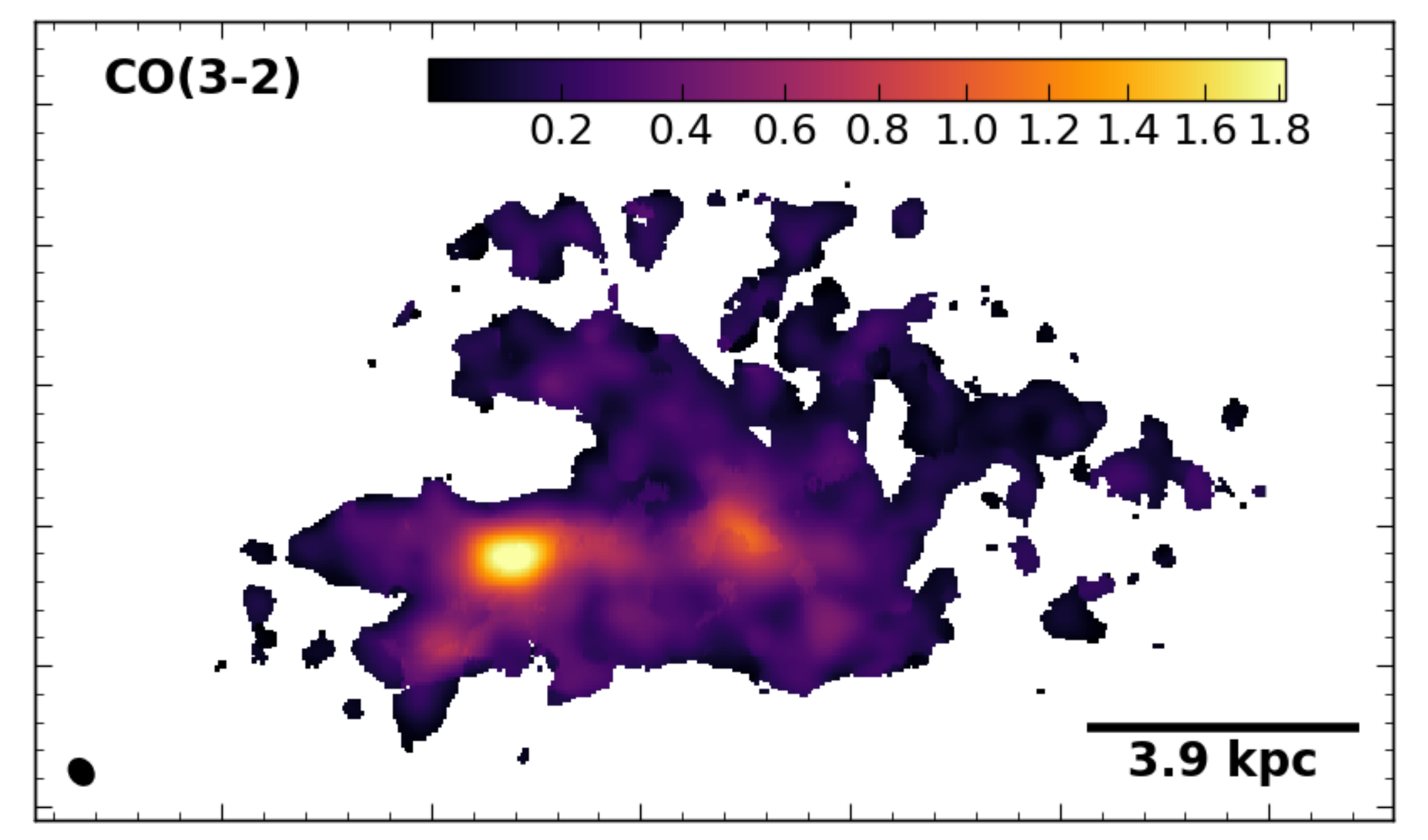}
  \caption[CO(1-0) and CO(3-2) total flux maps]{
    Maps of the CO(1-0) (top) and CO(3-2) (bottom) total fluxes, in $\Jykmps$. The $19.1\times 16.3\kpc$ field-of-view is the same in both images. The galactic center is not present in these images. The black ellipses in the lower left corners indicate the size of the synthesized beams. 
  }
  \label{fig:fluxmaps}
\end{figure}

Fig. \ref{fig:fluxmaps} presents the CO(1-0) and CO(3-2) integrated flux maps. These are updated versions of the flux maps presented in V17, using 2500 Monte Carlo iterations versus the 1000 iterations in V17. The fluxes of all velocity components have been summed together to create these maps. As discussed in V17, two bright clumps are situated along a $6\kpc$ long filament. The clumps account for more than half of the total line emission, and are surrounded by an envelope of diffuse emission. The diffuse emission extends to the NW in CO(1-0), while significant CO(3-2) emission is located north of the two clumps and does not extend as far to the NW.

All of this emission is spatially offset from the galactic center. The center of the brightest clump is $3\kpc$ north of the BCG's optical centroid. The other bright clump is $4.6\kpc$ NW of the optical centroid. The farthest extent of the filament is about $7\kpc$ from the galactic center.

In V17 we estimated that $<1.2\e{8}\Msun$ of molecular gas is present within a $2\times 2 \kpc$ box placed at the optical center of the BCG. This region is not located in the field-of-view of Fig. \ref{fig:fluxmaps}. Contours of the total CO(3-2) flux are overlaid on the optical emission in Fig. \ref{fig:xray} (right) for reference. Note that this mass limit was determined using the Galactic CO-to-H$_2$ conversion factor, which is double the value used in the rest of our mass estimates (see Section \ref{sec:mass}). We retained the Galactic conversion factor for this measurement in order to be conservative with the upper limit.

\begin{figure*}
  \centering
  \begin{minipage}{\textwidth}
    \centering
    \includegraphics[width=0.45\columnwidth]{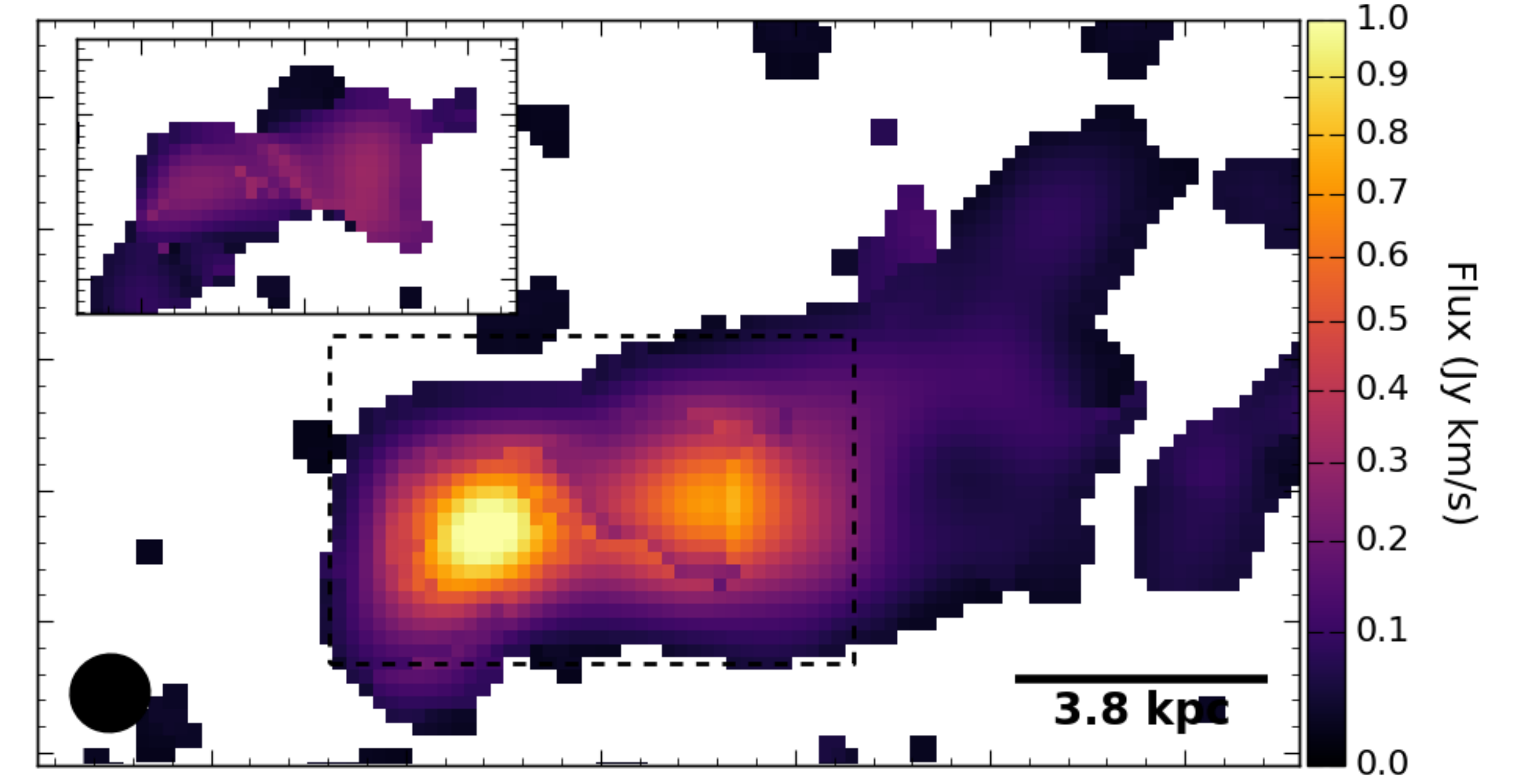}
    \includegraphics[width=0.45\columnwidth]{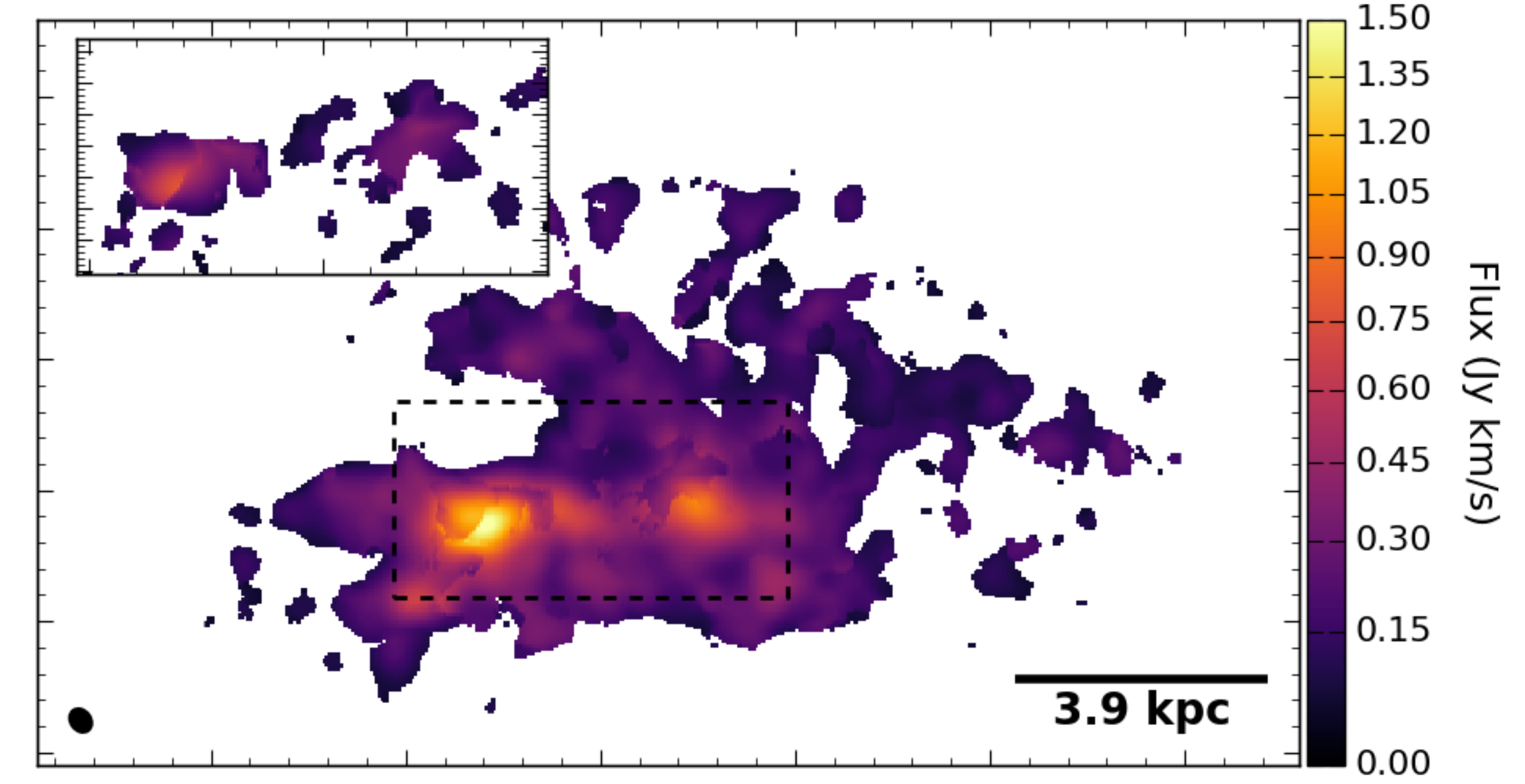}
  \end{minipage}
  \begin{minipage}{\textwidth}
    \centering
    \includegraphics[width=0.45\columnwidth]{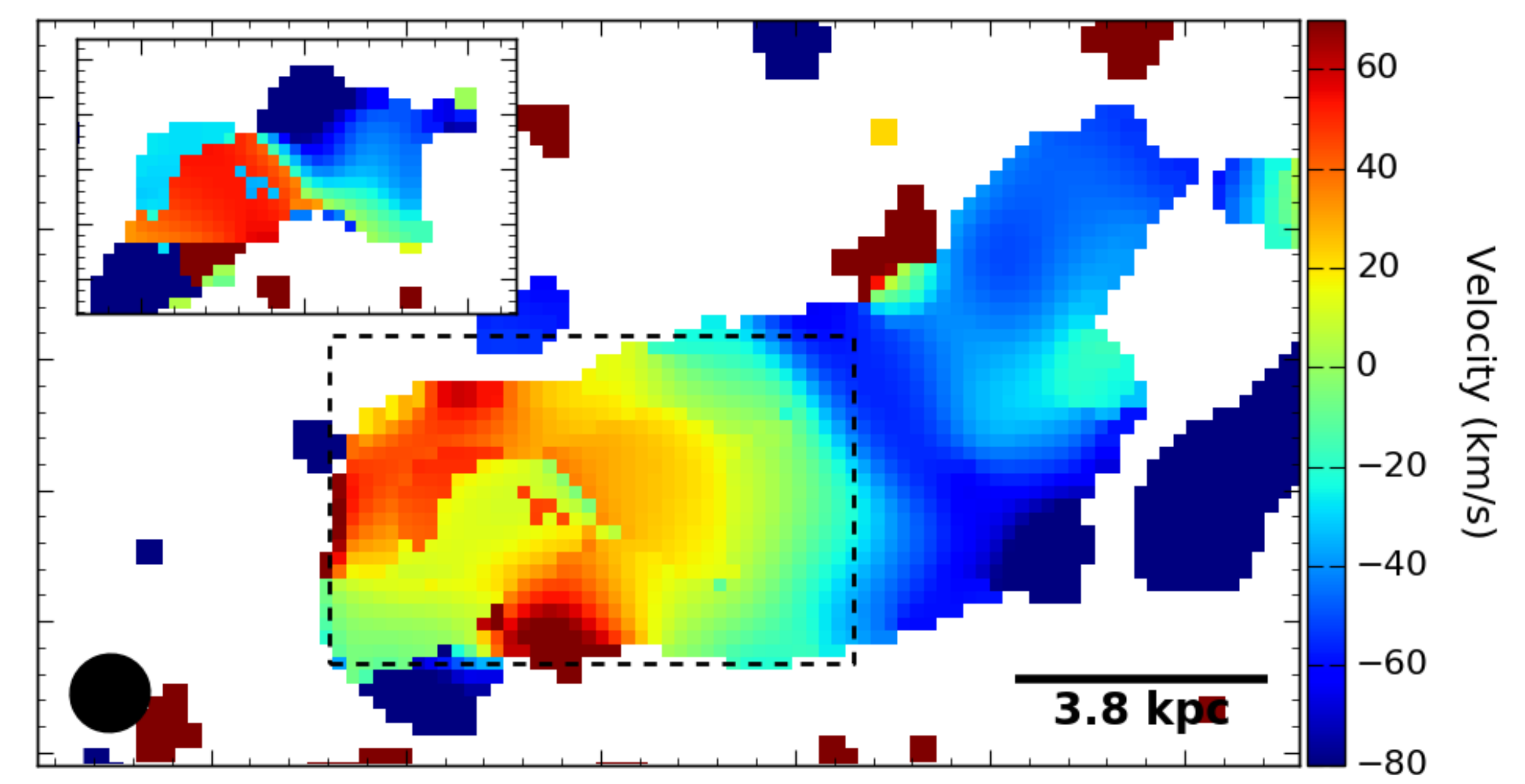}
    \includegraphics[width=0.45\columnwidth]{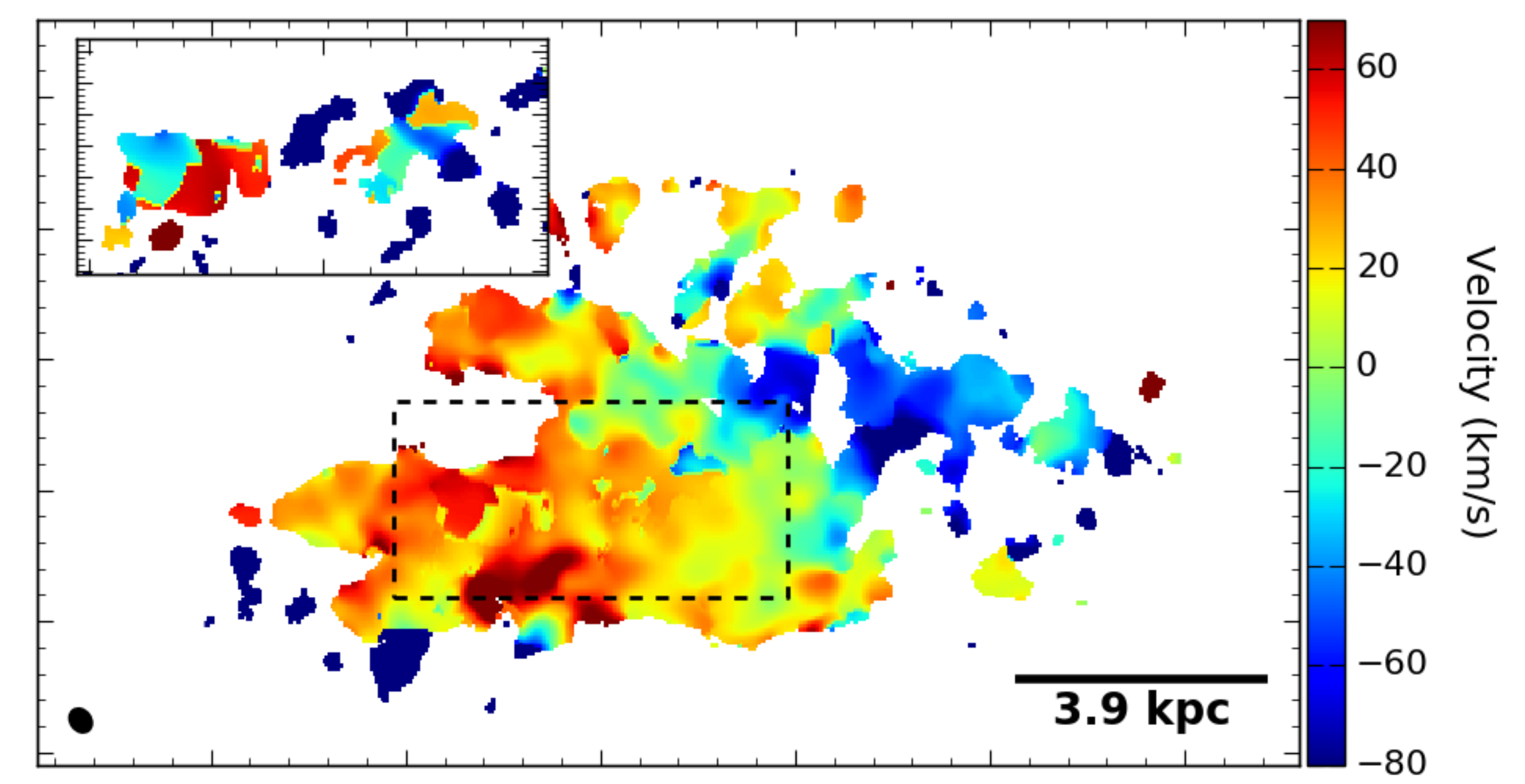}
  \end{minipage}
  \begin{minipage}{\textwidth}
    \centering
    \includegraphics[width=0.45\columnwidth]{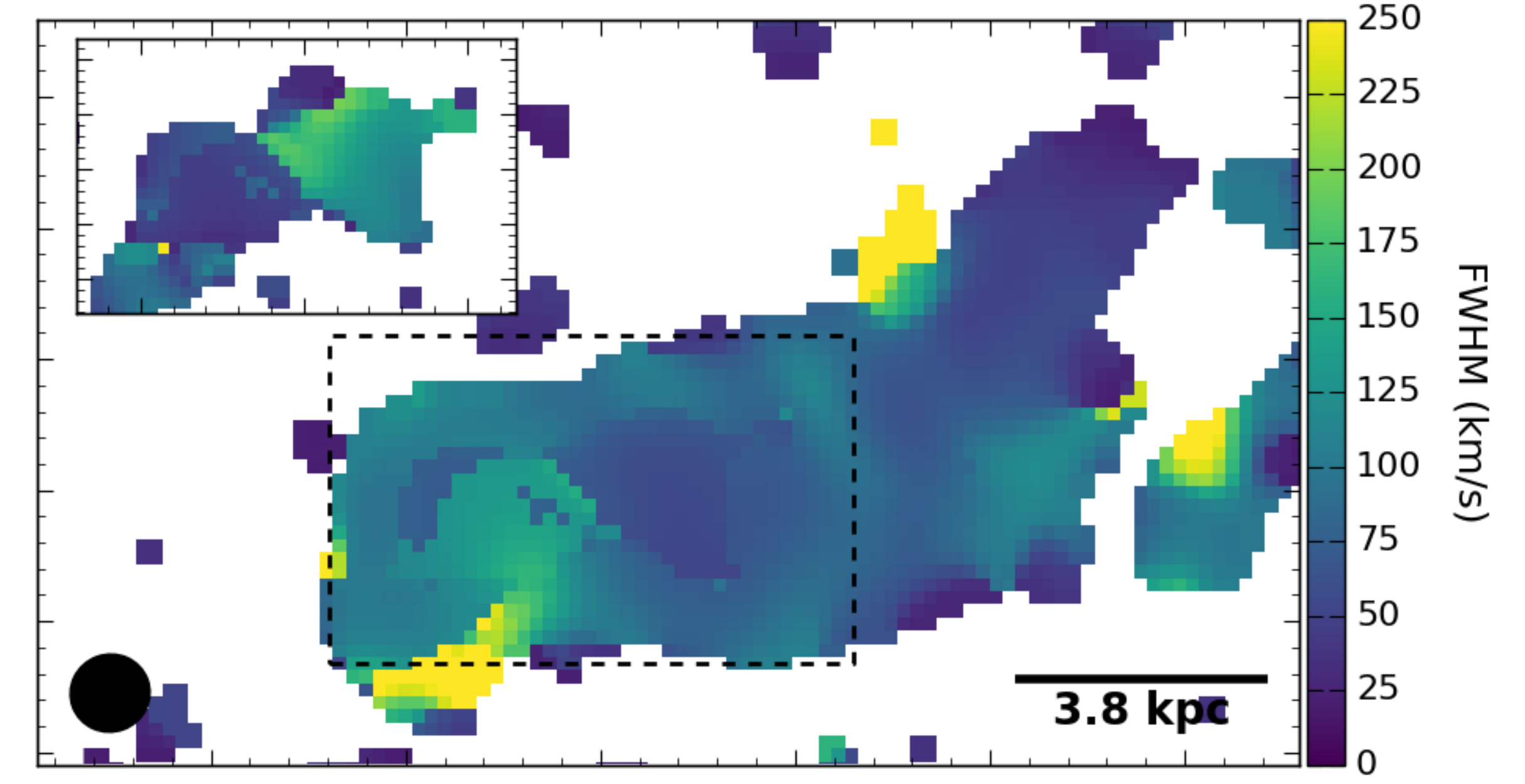}
    \includegraphics[width=0.45\columnwidth]{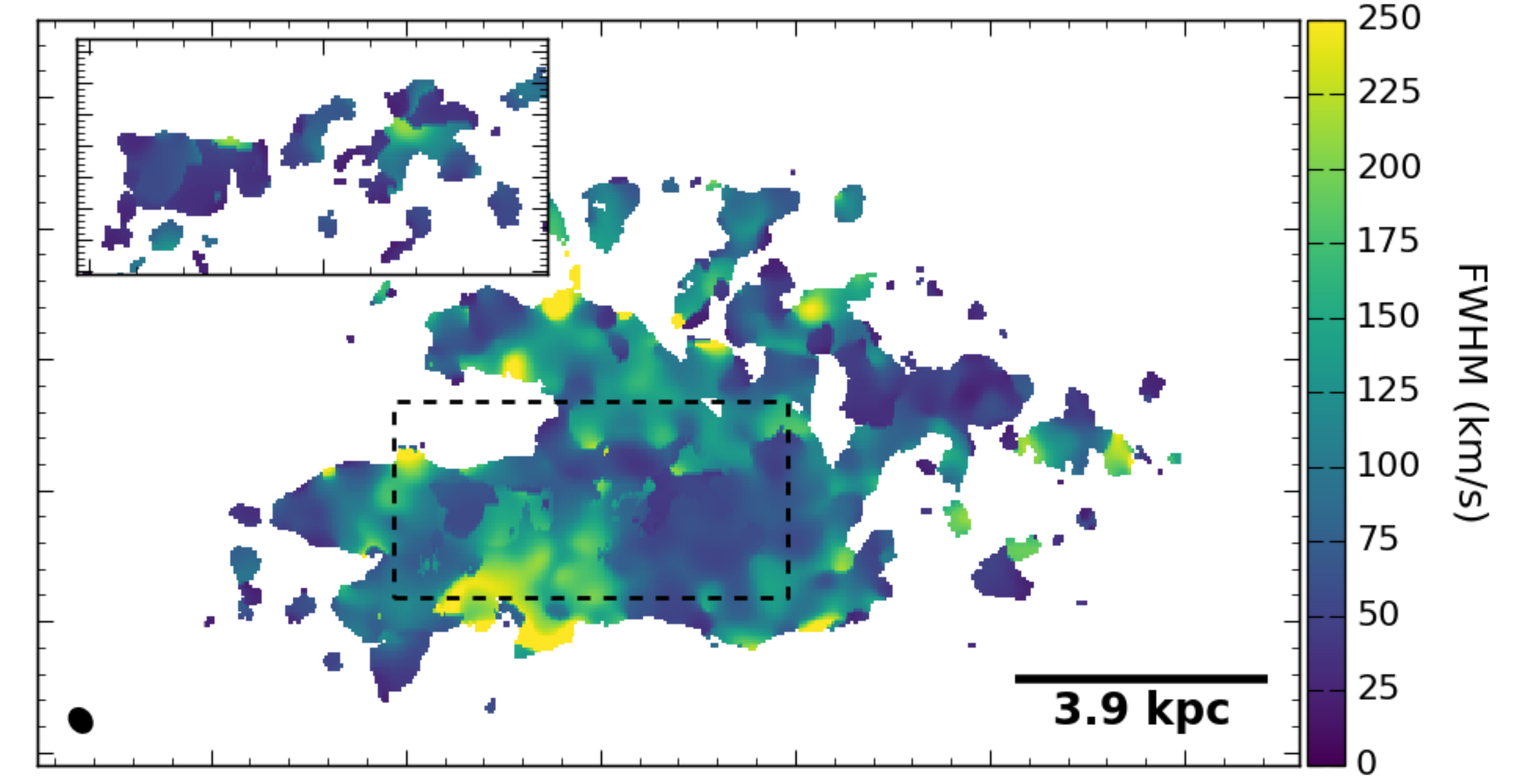}
  \end{minipage}
  \caption[CO(1-0) and CO(3-2) flux, velocity, and FWHM maps of RXJ0821]{Maps of the CO(1-0) (left) and CO(3-2) (right) integrated flux (top), velocity centroid (middle), and full-width at half maximum (bottom) for multi-component fits to each pixel. The field-of-view is the same as Fig. \ref{fig:fluxmaps}. The inset plot in the upper left corner shows the second velocity component within the region indicated by the dashed box.
  }
  \label{fig:velmaps}
\end{figure*}

The maps of integrated flux, velocity centroid, and FWHM are shown in Fig. \ref{fig:velmaps}. These maps are presented for both CO(1-0) (left) and CO(3-2) (right). For pixels whose spectra are best fit by two Gaussians, both velocity components are presented in Fig. \ref{fig:velmaps}. The main image is the velocity component with the largest flux, while the component with the lower flux is shown in the inset plot in the upper left corner. The dashed box indicates the region shown in the second plot. Significant detections of multiple velocity components were located near the two bright peaks. These inset regions measure $8\kpc\times 5\kpc$ in the CO(1-0) images and $6\kpc\times 3\kpc$ in the CO(3-2) images.

The molecular filament is separated into two regions of distinct velocities. 
The main portion of the molecular filament, containing the two bright clumps and approximately indicated by the dashed boxes in Fig. \ref{fig:velmaps}, exhibits a narrow range of velocities. The velocities span $0$ to $50\kmps$ throughout this region, with linewidths that are $<100\kmps$ FWHM. In the outer tail of the filament detected in CO(1-0) the velocity is blueshifted to $-25\kmps$. No coherent velocity structures are evident within either of these regions. The only coherent velocity gradient is in between these regions, where the velocity transitions sharply from $25\kmps$ to $-45\kmps$ over the span of $1.5\kpc$.

\subsection{Molecular Gas Mass}
\label{sec:mass}

The integrated flux ($S_{\rm CO}\Delta v$) of the CO(1-0) line can be converted to molecular gas mass through \citep{Solomon87, Solomon05, Bolatto13}
\begin{equation}
  M_{\rm mol} = 1.05\e{4} \frac{X_{\rm CO}}{X_{\rm CO, gal}} 
                \left(\frac{S_{\rm CO}\Delta v~D_L^2}{1+z}\right)\Msun.
  \label{eqn:Mmol}
\end{equation}
Here $z$ is the redshift of the source, $D_L$ is the luminosity distance in Mpc, and $S_{\rm CO}\Delta v$ is in $\Jykmps$. 

The CO-to-H$_2$ conversion factor within the Milky Way and other nearby spiral galaxies is measured to be $X_{\rm CO, gal} = 2\e{20}\pcmsq (\K\kmps)^{-1}$ \citep{Bolatto13}. Between similar systems $X_{\rm CO}$ varies by about a factor of two. In V17 we used the $^{13}$CO(3-2) emission line, in conjunction with the $^{12}$CO 1-0 and 3-2 lines, to estimate the conversion factor in RXJ0821. We measured $X_{\rm CO}=1.05\e{20}\pcmsq(\K\kmps)^{-1}$, or equivalently $\alpha_{\rm CO} = 2.26\acounits$. This is half the Galactic value. We have adopted this sub-Galactic value throughout this paper.

RXJ0821 is the only BCG for which a calibration of $X_{\rm CO}$ is available. This is an advantage over other systems, where the molecular gas mass may be overestimated. However, significant systematic uncertainties were unavoidable in the V17 analysis. In particular, the $^{13}$CO(3-2) measurement provides a measure of the $^{13}$CO column density, while N(H$_2$) is the desired quantity. This required the assumption of the $^{13}$CO/H$_2$ abundance ratio, which is only known to a factor of a few. The subsolar metallicity of the central intracluster medium in RXJ0821 suggests that $X_{\rm CO}$ may have been underestimated by a factor of 2--3, which would bring its value in line with the Galactic measurement.

In spite of these systematic uncertainties, we still adopt the sub-Galactic value of $X_{\rm CO}$. As a result, our reported molecular gas masses are conservative. Even these conservative masses, as discussed later, place stringent demands on the energetics in this system. Reverting to a Galactic conversion factor simply amounts to multiplying the molecular gas masses by a factor of two.

Following V17, CO(3-2) line fluxes have been converted to molecular gas masses by assuming a CO (3-2)/(1-0) flux ratio of $8$. This was measured from the ratio of integrated flux densities. Using the ratio of peak temperatures instead would give a lower line ratio of $6.4$. 

For total integrated CO(1-0) and CO(3-2) fluxes of $8.06\pm0.08$ and $65.6\pm1.5\Jykmps$, respectively, the molecular gas mass is $(1.07\pm0.02)\e{10}\Msun$.

\begin{figure}
  \centering
  \includegraphics[width=\columnwidth]{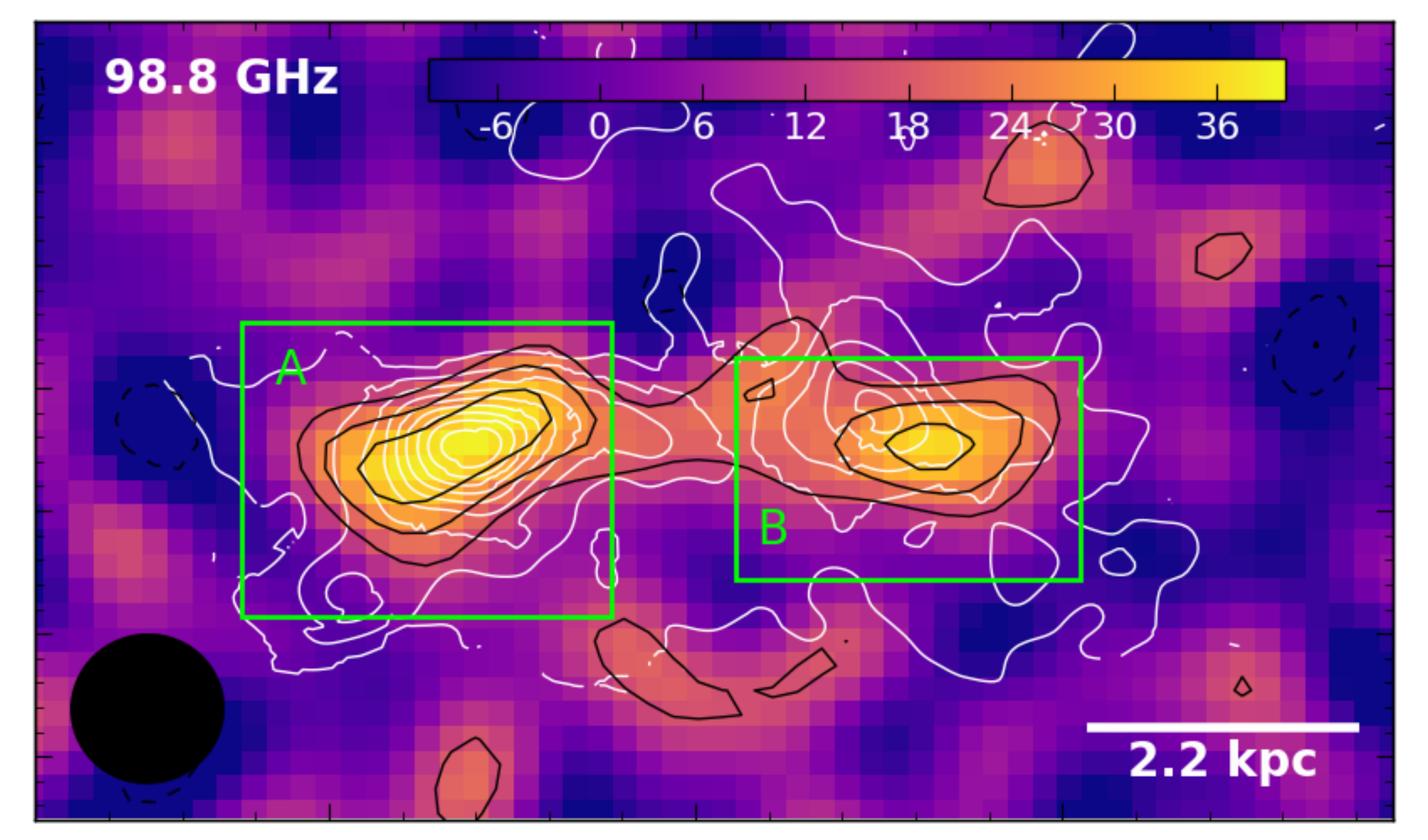}
  \includegraphics[width=\columnwidth]{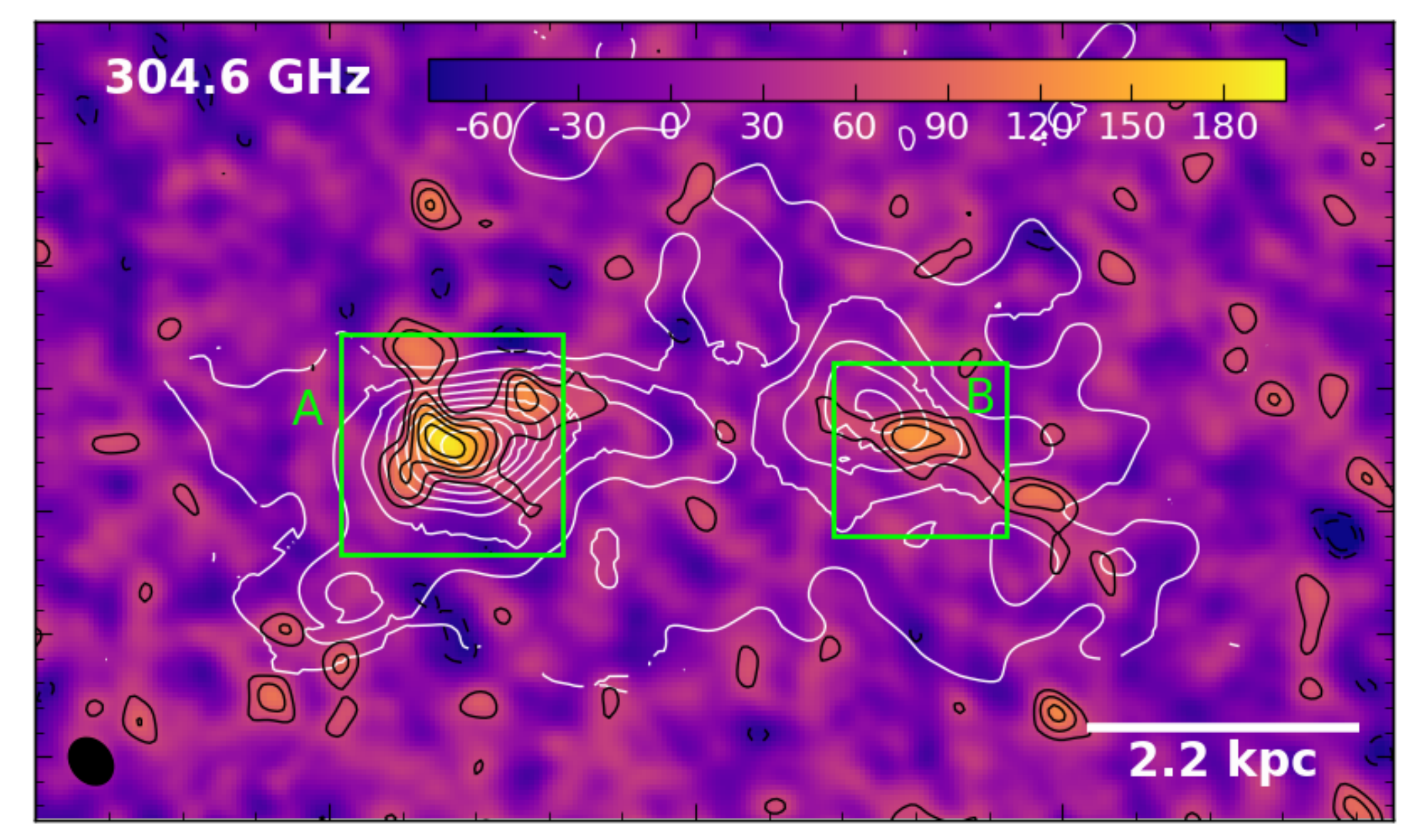}
  \caption[RXJ0821 dust continuum]{
    The $98.6\GHz$ (top) and $304.6\GHz$ (bottom) continuum sources. The continuum flux is traced by black contours at the [-3, -2, 2, 3, 4, 5, 6]$\times\sigma$ levels, where $\sigma=8.2\uJy~{\rm beam}^{-1}$ at $98.6\GHz$ and $28\uJy~{\rm beam}^{-1}$ at $304.6\GHz$. The colorbar is in units of $\uJy~{\rm beam}^{-1}$. The white contours correspond to the ALMA CO(3-2) flux (Fig. \ref{fig:fluxmaps} bottom). The continuum measurements provided in Table \ref{tab:continuum} were extracted from the regions shown in green.
  }
  \label{fig:continuum}
\end{figure}

\section{Dust Continuum}
\label{sec:continuum}

Continuum maps from ALMA Bands 3 ($98.8\GHz$) and 7 ($304.6\GHz$) are presented in Fig. \ref{fig:continuum}. No radio continuum is detected at the BCG nucleus. Instead, very faint, extended emission located near the two main clumps of molecular gas are present at both frequencies. The faint continuum emission in Fig. \ref{fig:continuum} is highlighted using black contours at uniform intervals of $\sigma$, where $\sigma$ is $8.2\uJy~{\rm beam}^{-1}$ at $98.6\GHz$ and $28\uJy~{\rm beam}^{-1}$ at $304.6\GHz$. White contours indicate the integrated CO(3-2) line emission from Fig. \ref{fig:fluxmaps}.

Measurements of the continuum flux density for both clumps (labelled as regions A and B in Fig. \ref{fig:continuum}) are presented in Table \ref{tab:continuum}. The box sizes used to extract the fluxes were chosen based on the size of each source, so differ between the two frequencies because of the differing resolutions. Each box was centered on the source and grown until the signal-to-noise fell to $1.5$. Since the emission is extended, some flux is missing from the adopted regions and our measurements should be considered lower limits. Although each individual measurement is marginal, the combination of all four independent measurements corresponds to a $3.5\sigma$ detection of the continuum.

RXJ0821 is the third brightest {\it Spitzer} $70\mum$ source in a sample of 62 BCGs \citep{Quillen08}. Its prominent red unresolved nucleus at $8\mum$ and high [O~{\sc iii}](5007)/H$\beta$ ratio suggests that it hosts a dusty AGN. The high IR flux and spatial coincidence with the molecular gas implies that the ALMA continuum also originates from dust emission. This is the best example of resolved dust continuum from ALMA.

\begin{table}
\caption{Radio Continuum}
\begin{center}
\begin{tabular}{l c c c}
\hline\hline
Region & Frequency & Region Dimensions & Flux Density \\
       & (GHz)     & (kpc$\times$kpc)  & (mJy)       \\
\hline
A   & $98.8$    & $3.0\times 2.4$ & $0.066\pm0.043$ \\
    & $304.6$   & $1.8\times 1.8$ & $0.90\pm0.60$ \\
B   & $98.8$    & $2.8\times 1.8$ & $0.05\pm0.03$ \\
    & $304.6$   & $1.4\times 1.4$ & $0.51\pm0.37$ \\
\hline\hline
\end{tabular}
\end{center}
\label{tab:continuum}
\end{table}

\section{Discussion}

RXJ0821 presents an interesting challenge to our understanding of molecular gas formation and flows in BCGs. All of its $10^{10}\Msun$ of molecular gas is offset by $\sim4\kpc$ from the galactic center. This is among the most massive known molecular gas reservoirs in a BCG, and evidently none of it has settled into the underlying gravitational potential. Additionally, the narrow spread in both position and velocity indicates that the molecular gas has either formed rapidly, having had little time for infall, or it has been deposited abruptly.

Observations and simulations consistently support the hypothesis that ICM condensation is the primary source of cold gas in galaxy clusters \citep[e.g.][]{Rafferty08, McDonald10, Gaspari13, Li14a, Russell16, Vantyghem16, Pulido18}. Most directly, the presence of cold gas and star formation is linked to short central cooling times \citep{Cavagnolo08, Rafferty08, Pulido18}. 
Mergers are, in general, unable to account for the massive molecular reservoirs observed in BCGs. Even gas-rich spirals, such as the Milky Way, contain less gas than is present in typical BCGs. Merger rates are also unrelated to the presence of a cool core, so the cooling time threshold cannot be explained through mergers.

\subsection{Gas Donated or Displaced by an Infalling Galaxy}

Despite the inability of mergers to form the molecular gas reservoirs of BCGs in general, the proximity of a nearby galaxy raises the possibility of a merger origin for the cold gas in RXJ0821. The galaxy SDSS J082102.46+075145.0 is located $7.7\kpc$ SE of the BCG nucleus. Optical spectroscopy indicates that its relative velocity is $+77\pm32\kmps$ with respect to the BCG \citep{BayerKim02}. Blue emission from recent star formation connects the molecular gas reservoir to the nearby galaxy. These factors suggest that the two galaxies have either interacted in the past or are currently interacting.


A gas-rich elliptical can contribute at most a few $\times 10^{8}\Msun$ of cold gas \citep{Young11}. Tens to hundreds of merging ellipticals would be required to accumulate the $10^{10}\Msun$ gas supply in RXJ0821. 

Alternatively, the nearby galaxy could be the remnant bulge of an infalling spiral. However the Milky Way -- a gas-rich spiral galaxy -- contains only $10^{9}\Msun$ of molecular gas \citep{Heyer15}. The nearby galaxy is potentially a few times more massive than the Milky Way. Its SDSS magnitudes, corrected for evolution and the K-correction \citep{Poggianti97}, yield an absolute $i$-band luminosity of $L_i = 1.35\e{10}\Lsun$. This includes a 25\% correction for the underlying BCG flux, determined using a pair (source and background) of adjacent $1''$ radius apertures in the HST F606W image. For an $i$-band mass-to-light ratio of $2.0$ \citep{Bell03}, the total stellar mass is $2.7\e{10}\Msun$. This is three times greater than the mass of the Milky Way's bulge \citep[0.91\e{10}\Msun;][]{Lacquia15a}. The same increase in molecular gas mass would still be three times lower than the molecular gas mass of RXJ0821.

Atomic gas provides another avenue for producing the massive molecular gas reservoir. The pressure in cluster cores is high enough to convert virtually all atomic gas to molecular form \citep{Blitz06}. The Milky Way contains $\sim6\e{9}\Msun$ of H\,{\sc i} \citep{Ferriere01}, and the nearby galaxy may have initially contained a few times more. The combination of the pre-existing molecular gas with the condensing H\,{\sc i} could conceivably account for the $10^{10}\Msun$ of cold gas in RXJ0821. 

In order to strip molecular gas the galaxy must be infalling with a high velocity. Even large spirals containing $10^{8}\Msun$ of molecular gas must be moving at $1000\kmps$ in order for their cold gas to be stripped $10\kpc$ from cluster cores \citep{Kirkpatrick09}. Larger galaxies would require even faster velocities. The low relative velocity of $77\kmps$ indicates that, unless virtually all of its motion is along the plane of the sky, the nearby galaxy should have held onto all of its molecular gas. 

H\,{\sc i} is stripped much more easily than H$_2$, as its density is $2-3$ decades lower. It is unlikely that H\,{\sc i} has survived long enough to condense into molecular gas. Indeed, spirals in cluster environments are deficient in H\,{\sc i} \citep{Haynes84}. Additionally, ram pressure tails differ morphologically from the gas in RXJ0821. When velocities are high enough to strip molecular gas, the resulting tail can extend tens of kpc outward from the infalling galaxy, occupying a wide range in both position and velocity \citep[e.g.][]{Jachym14, Jachym17}. 

Another possibility is that the $10^{10}\Msun$ of molecular gas was initially central within the BCG but was dislodged by the infalling galaxy. Rings and partial rings of gas and young stars are observed following nearly head-on collisions with infalling galaxies \citep{Appleton96}. The arcs of blue emission and molecular gas could be a partial ring driven outward by the interaction with the nearby galaxy. However, several lines of reasoning lead us to argue that this is not the case. 

First, the molecular gas is coincident with an X-ray-bright plume (discussed further in Section \ref{sec:condensation}). This correlation implicates ICM condensation as a formation mechanism and would not be caused by a minor merger. Next, these collisions do not remove the entire gas supply \citep{Lynds76}. Less than $1\%$ of the molecular gas in RXJ0821 resides at the galactic center, so any mechanism that displaces the gas must be efficient. Moreover, the molecular gas occupies a narrow range in both space and velocity, which is inconsistent with a high speed collision. Finally, an infalling galaxy on its first passage through the cluster center would have a velocity in excess of $1000\kmps$. The observed radial velocity of $+77\kmps$ implies that the nearby galaxy would be travelling within $5^{\circ}$ of the plane of the sky. This is possible, but statistically unlikely.


Overall the possibility that the cold gas has been either deposited or dislodged by a merger is not well-motivated by the observational data.

\subsection{ICM Condensation}
\label{sec:condensation}

The spatial coincidence between molecular gas and the X-ray-bright plume (see Fig. \ref{fig:xray}) supports the possibility that the cold gas has condensed out of the hot atmosphere. ICM condensation is easiest in cluster cores, where the cooling time is shortest. When radiative cooling is approximately balanced by AGN heating condensation ensues via thermal instabilities \citep[e.g.][]{McCourt12, Gaspari13, Li14b, Voit18}. This requires the central gas with a short cooling time to be displaced from its equilibrium position long enough for the gas to cool. Two ways to accomplish this are uplift in the wakes of X-ray cavities and sloshing of the ICM.

Before discussing either of these possibilities, it is important to note that even condensation of the hot atmosphere has difficulty accounting for such a massive reservoir of molecular gas. 
The hot gas mass within the inner region ($13.3\kpc$ radius) of the X-ray profiles (Fig. \ref{fig:profiles}) contains $(1.27\pm0.14)\e{10}\Msun$, which is comparable to the molecular gas mass. The narrow spatial and velocity distributions indicate that the molecular gas formed rapidly and in a single cycle of cooling. Rapid condensation would deplete the central $10\kpc$ of its entire supply of hot gas, resulting in a rapid inflow to balance the pressure support lost as the gas condenses. This is not unprecedented. Eight of the 33 systems in \citet{Pulido18} that contain significant CO emission have molecular gas masses that match or exceed the hot gas mass within $10\kpc$. 

Additionally, the mass deposition rate within the central $50\kpc$ is $<34\pm10\Msunpyr$ (see Section \ref{sec:cooling}). Condensation persisting at this rate would form the $10^{10}\Msun$ of molecular gas in $3\e{8}\yr$. This is close to the central cooling time of $4\e{8}\yr$. However, star formation is also present in the BCG. Infrared measurements imply a star formation rate of $37\Msunpyr$ \citep{ODea08}. Condensation at this rate should therefore be largely offset by star formation, resulting in a slowly accumulating gas reservoir.

Non-radiative ICM cooling may help, but not completely, alleviate these demands. Heat transfer with the molecular gas (e.g. conduction, collisions, mixing) can hasten the overall rate of cooling. This would make it easier for a single cycle of cooling to produce the molecular flow, but still suffers from the lack of fuel in the central $10\kpc$.

\subsubsection{Stimulated Cooling}

Stimulated cooling is emerging as a leading mechanism in triggering the formation of molecular gas in BCGs. In this mechanism, low entropy gas from the cluster core is lifted by rising X-ray cavities to an altitude where it becomes thermally unstable \citep{Revaz08, mcn16}. Stimulated cooling was proposed in response to the growing number of ALMA observations with molecular filaments trailing X-ray cavities. The molecular gas in RXJ0821 also exhibits a connection with an X-ray cavity, as it is coincident with the bright X-ray plume that wraps around the northern side of the cavity.

ALMA observations of other BCGs have demonstrated that X-ray cavities are capable of lifting enough low entropy gas to form their trailing molecular filaments, although the coupling efficiencies must be high \citep[e.g.][]{mcn14, Russell16, Russell17, Russell17b, Vantyghem16}. In RXJ0821, on the other hand, the cavity is far too feeble to have lifted $10^{10}\Msun$ of gas. Archimedes' principle provides a convenient metric to explore the feasibility of uplift behind a rising cavity. Cavities cannot lift more gas than they displace. The mass of the displaced ICM is given by $M_{\rm displaced} = n\mu m_H V$, where $n=n_e+n_H$ is the density of the surrounding gas. A total of $2.4\e{8}\Msun$ of hot gas has been displaced in inflating the X-ray cavity. This is 40 times smaller than the molecular gas mass. Only a few percent of the total molecular gas mass could have been uplifted by the cavity. While uplift behind X-ray cavities is a promising cold gas formation mechanism in other BCGs, it fails here.

The morphology of RXJ0821's molecular filament also differs somewhat from the filaments in other BCGs. Half of the gas in RXJ0821 is concentrated in two clumps surrounded by a diffuse, $2-5\kpc$ wide envelope. Molecular filaments in other systems vary in length from $\sim3-20\kpc$ with unresolved widths that are $\lesssim 1\kpc$ \citep[e.g.][]{Russell16, Russell17, Russell17b, Vantyghem16, Vantyghem18}. Clumpy emission is also observed in BCGs \citep[e.g.][]{mcn14, Tremblay16}, but is often poorly resolved and coincident with the galactic center. The somewhat unique structure in RXJ0821 may be indicative of a different formation mechanism.

\subsubsection{Sloshing}

Condensation could also be triggered by sloshing motions in the ICM. Minor mergers can easily set the central peak of the ICM in motion with respect to the rest of the cluster \citep{Ascasibar06}. These sloshing motions persist for several Gyr, potentially providing enough time for the central gas to cool. The low relative velocity and arcs of blue continua trailing the nearby galaxy suggest that a minor merger has occurred. Further indications of sloshing come from the mutual offsets between the X-ray peak, BCG nucleus, and the centroid of the X-ray emission on $20\kpc$ scales. The molecular gas is offset from each of these, but is coincident with the bright plume that extends from the X-ray peak.

Sloshing may also contribute to the formation of cold gas in other systems. A1795 hosts a $50\kpc$ long filament extending from the BCG \citep{Fabian01, McDonald09, McDonald12b}. \citet{Fabian01} argued that this filament was formed via gravitational focusing as the BCG passed through the ICM. \citet{Hamer12} identified three clusters (A1991, A3444, and Ophiuchus) with X-ray peaks offset from the BCG by $\sim10\kpc$. The nebular emission and molecular gas in these systems are coincident with the soft X-ray peak, implying a causal link between the lowest temperature ICM and the molecular gas. In each case the large offset between the BCG and X-ray peak is attributed to major or minor cluster mergers. The smaller offset in RXJ0821 requires a smaller perturbation. 

Whether sloshing is able to account for all $10^{10}\Msun$ of cold gas in RXJ0821 is unclear. It would still require the condensation of all of the ICM within the central $10\kpc$, but this gas is already oscillating within the cluster potential. In stimulated cooling X-ray cavities would need to couple to and lift the same amount out of the cluster core, which is much more difficult. We therefore argue that sloshing is the mechanism responsible for triggering condensation in RXJ0821.

\section{Conclusions}

In this work we have performed a morphological analysis of ALMA CO(1-0) and CO(3-2) observations and presented a new $63.5\ks$ {\it Chandra} X-ray observation of the cool core cluster RXJ0821.0+0752. This extends the previous analysis of the same ALMA data conducted in V17, where the analysis focused on using the line intensities, along with a $^{13}$CO(3-2) detection, to constrain the CO-to-H$_2$ conversion factor.

The entire $10^{10}\Msun$ supply of molecular gas is located in an $8\kpc$ long filament spatially offset from the galactic center by about $4\kpc$. The emission is concentrated in two bright peaks surrounded by a diffuse, $2-5\kpc$ wide envelope. It is coincident with a bright plume of X-ray emission situated alongside a putative X-ray cavity. The narrow spread in both position and velocity suggests that the molecular gas formed relatively quickly and in a single cycle of cooling.

The formation of cold gas in RXJ0821 differs from that of other BCGs, where stimulated cooling has lifted filaments of condensing gas out of the cluster core. Although the filament in RXJ0821 is also associated with an X-ray cavity, the cavity is too feeble to account for the observed gas distribution. Only a few percent of the total molecular gas mass could have been uplifted by the cavity.

Instead, ICM condensation in RXJ0821 has likely been triggered by sloshing motions induced by the interaction with a nearby galaxy. The BCG nucleus, X-ray peak, molecular gas, and arcs of recent star formation are all mutually offset, which is indicative of relative motions between the cooling ICM and BCG. Sloshing can trigger condensation by emulating uplift. The cooling time is shortest in the cluster core. Sloshing removes this gas from the center of the potential well, giving some of the gas time to cool by keeping it out of equilibrium.

\acknowledgements

Support for this work was provided in part by the National Aeronautics and Space Administration through Chandra Award Number G08-19109A issued by the Chandra X-ray Observatory Center, which is operated by the Smithsonian Astrophysical Observatory for and on behalf of the National Aeronautics Space Administration under contract NAS8-03060.
BRM acknowledges support from the Natural Sciences and Engineering Research Council of Canada.
BRM further acknowledges support from the Canadian Space Agency Space Science Enhancement Program.
ACE acknowledges support from STFC grant ST/P00541/1.
This paper makes use of the ALMA data ADS/JAO.ALMA 2016.1.01269.S. ALMA is a partnership of the ESO (representing its member states), NSF (USA) and NINS (Japan), together with NRC (Canada), NSC and ASIAA (Taiwan), and KASI (Republic of Korea), in cooperation with the Republic of Chile. The Joint ALMA Observatory is operated by ESO, AUI/NRAO, and NAOJ.
This research made use of Astropy, a community-developed core Python package for Astronomy.
This research made use of APLpy, an open-source plotting package for Python hosted at http://aplpy.github.com.

\bibliographystyle{apj}
\bibliography{rxj08}

\end{document}